\documentclass[a4paper,10pt]{article}
\usepackage{amsmath}
\usepackage{graphicx}
\usepackage{subfigure}
\usepackage{fancyhdr}
\usepackage{a4wide}
\newcommand{\refer}[1]{(\ref{#1})}
\DeclareMathOperator{\im}{Im}

\def\tempOp#1{\ensuremath{O\!\left(p^{#1}\right)}}
\def\Op#1{\protect\tempOp#1}
\def\Opl#1{\ensuremath{O\!\left((p/\Lambda)^{#1}\right)}}

\def\Ka{{\overline{K^0}}}

\def\chiPT{$\chi\textrm{PT}$}
\def\Delt#1{\ensuremath{\Delta_{\scriptscriptstyle #1}}}

\def\jednotka#1{\ensuremath{\,\mathrm{#1}}}

\def\caligr#1{\ensuremath{\mathcal{#1}}}
\def\M{\caligr{M}}

\def\bracketif<#1|#2>{\ensuremath{\left\langle #1 \vert_{f} #2\right\rangle_{i}}}

\title{Dispersive Approach to Chiral Perturbation Theory}
\author{M.\ Zdr\'{a}hal$^{a,b,}$\footnote{zdrahal@ipnp.troja.mff.cuni.cz}, J.\ Novotn\'{y}$^{a,}$\footnote{novotny@ipnp.troja.mff.cuni.cz}\\
\\
\small $^a$Institute of Particle and Nuclear Physics, Faculty of Mathematics and Physics,\\
\small Charles University, V Hole\v{s}ovi\v{c}k\'{a}ch 2, CZ-180
00 Prague 8, Czech Republic\\
\small $^b$Faculty of Physics, University of Vienna,
\\
\small Boltzmanngasse 5, A-1090 Vienna, Austria}
\begin{document}
\maketitle

\thispagestyle{fancy}
\renewcommand{\headrulewidth}{0.0pt}
\lhead{}
\chead{}
\rhead{UWThPh-2008-10}
\lfoot{}
\cfoot{}
\rfoot{}

\begin{abstract}
We generalise the reconstruction theorem of Stern, Sazdjian,
and Fuchs based on the dispersion relations to the case of the
($2\rightarrow2$) scattering of all the pseudoscalar octet mesons
($\pi, K, \eta$). We formulate it in a general way and include also a discussion of the assumptions of the theorem.
It is used to obtain the amplitudes of all such processes in the isospin
limit to the one-loop order (and can be straightforwardly extended
to two loops) independently on the particular power-counting scheme
of the chiral perturbation theory in question. The results in this
general form are presented.
\end{abstract}

\section{Introduction}
In low-energy QCD the chiral perturbation theory (\chiPT,
cf.\ \cite{review}) has gained great prominence as a tool for
description interactions of the lightest (pseudo\-scalar) mesons,
$\pi$, $K$ and $\eta$. Within this framework these mesons are
understood as Goldstone bosons of the spontaneous breaking of the
chiral symmetry $SU(3)_L\times SU(3)_R$ (appearing if all three
lightest quarks were massless) down to $SU(3)_V$, gaining the masses
through the explicit breaking of the symmetry due to the nonzero
quark masses, whereas all the other hadrons (with mass at least of order
$\Lambda\sim1\jednotka{GeV}$) are included only effectively. Using
the symmetry properties and the basic properties like analyticity,
unitarity, and the crossing symmetry, the Lagrangian of this effective
theory is constructed. It contains an infinite number of terms.
Nevertheless, the Weinberg power-counting scheme assigns to a given
diagram of this theory (and term of the Lagrangian) its importance
by means of the chiral dimension \Op{D} (see \cite{Weinberg}), and so
for computations to a given order in low energies it is sufficient
to use only a finite number of them.

However, we can try to find the amplitudes of a process using the
required properties directly with no need of the Lagrangian to be
explicitly written (with all the advantages and disadvantages that this
involves), which is the main goal of this paper. We extend the
reconstruction theorem given by Stern, Sazdjian, and Fuchs in
\cite{SSF} and widely used in \cite{SSF2}. Their work was originally
motivated by the problem\footnote{The standard power counting is
based on the presumption that the value of $B_0$ is of order $\Lambda$.
So there also appeared approaches extending the \chiPT\ to the case
including the possibility of small $B_0$ too, which have a lesser
predictive power connected with larger number of LEC - namely the
generalised \chiPT\ \cite{Generalised}.} of possible smallness of
the scalar condensate $B_0$ as the power-counting independent way
to compute the amplitude of $\pi\pi$ scattering in the 2 flavour
case. Even though the problem is not yet satisfactorily
closed\footnote{The only direct answer given by experiment is still
the confirmation of the standard power counting in the 2 flavour
case by the $K_{l4}$ measurement at BNL \cite{Kl4}.}, the method is
interesting also in itself. The extension of this method to the 3
flavour case of the $\pi K$ amplitude has been given by
Ananthanarayan and B\"{u}ttiker \cite{Moussalamm piK}. We have
worked out a generalisation to the processes of all the pseudo-Goldstone bosons (pGB) in the
three flavour case.

In contrast to these previous papers, we have extracted from the
theorem the entire information about the isospin structure and
assumed just the needed crossing symmetry. The isospin properties
are used afterwards separately for particular processes.
Furthermore, the theorem is formulated formally generally by
pointing out all the properties essential for the theorem. This
should simplify the discussion of validity of the theorem and its
eventual further extensions to more complicated cases. In the simple
case of processes from the pure strong \chiPT\ (conserving the strong
isospin), we have discussed the assumptions and showed their
reasonability and concluded with the construction of their
amplitudes to two loops (with results explicitly written just to the
one loop) in a way independent of the particular power-counting scheme
in use. In this aspect our work is thus also an extension of the work of Osborn \cite{Osborn}
to higher orders.

The plan of the paper is as follows. In Sec.~2 we establish
our notation; the fundamental content of the paper is Sec.~3, where
we reformulate the reconstruction theorem with its assumptions and
give the proof of it. In Sec.~4 we use the theorem (and Appendix~A) to compute the amplitudes describing all the 2 pGB
scattering processes to the \Op4 chiral order (explicitly written in Appendix~B). Finally, in Sec.~5 we provide a further
discussion of the assumptions and discussion of the results.
Appendix~A summarises all the needed amplitudes to describe the
4-pGB processes and lists a general parametrisation of them to the
\Op2 order. Appendix~B gives the results of the application of
the theorem to computation of amplitudes to the \Op4 order.
In Appendix~C we give relations between our parameters and
renormalised LEC of the standard \chiPT\ \cite{Gasser} in
a particular renormalisation scheme \cite{Spanele}. The
main points of the analysis of validity of dispersion relations and
the regions of analyticity of amplitudes that are used in the
theorem are briefly summarised in Appendix~D.

\section{Notation}
We consider a scattering process of two pGB of the type
\begin{equation}
A(p_A)B(p_B)\rightarrow C(p_C)D(p_D)\label{process ABCD}
\end{equation}
and write its amplitude according to\footnote{This process depends
on two independent kinematical quantities only. It is convenient to
choose the total energy squared $s$ and the angle $\theta$ between
the momenta $p_A$ and $p_C$ in the center of mass system (CMS) or
two of three Mandelstam variables $s, t, u$. In order to express the crossing and Bose symmetry properties
in a simple way in what follows we keep writing the dependence on all three Mandelstam variables explicitly.}
\begin{equation}\label{definice amplitudy}
\bracketif<C(p_C)D(p_D)|A(p_A)B(p_B)>=\delta_{if}+i(2\pi)^4\delta^{(4)}(p_C+p_D-p_A-p_B)A(s,t,u).
\end{equation}
In the relations where we want to distinguish between more processes,
we write explicitly $A_{i\rightarrow f}(s,t,u)$ or $A_{AB\rightarrow
CD}(s,t,u)$.

Later on, we will use the crossing symmetry, so let us define the
amplitude of the crossed processes. The amplitude of the direct
process will be denoted by $S(s,t,u)$, i.e.\ $A_{AB\rightarrow
CD}(s,t,u)=S(s,t,u)$. The amplitude of the $T$-crossed channel will be
$A_{A\overline{C}\rightarrow \overline{B}D}(s,t,u)=f_T T(s,t,u)$,
where the phase factor $f_T$ is defined so that $T(s,t,u)$ fulfils
the crossing relation $S(s,t,u)=T(t,s,u)$ and similarly for the
$U$-crossed channel.

The transition from the CMS variables $s$ and $\cos\theta$ to the
Mandelstam variables $s,t,u$ for the process \refer{process ABCD} is
possible using the relation
\begin{equation}\label{uhel}
\cos\theta=\frac{s(t-u)+\Delt{AB}\Delt{CD}}{\lambda_{AB}^{1/2}(s)\lambda_{CD}^{1/2}(s)}\,,
\end{equation}
where $\Delt{ij}$ is a difference of the masses squared $\Delt{ij}=m_i^2-m_j^2$ and
\begin{equation}\label{triangle}
\lambda_{ij}(x)=\lambda(x,m_i^2,m_j^2)=x^2+m_i^4+m_j^4-2x(m_i^2+m_j^2)-2m_i^2m_j^2
\end{equation}
refers to the K\"{a}llen's quadratic form (also called the triangle
function).

The sum of squared masses of all the particles appearing in the
process is denoted as $\M$, i.e.\
\begin{equation}\label{soucet Mandelstamu}
s+t+u=m_A^2+m_B^2+m_C^2+m_D^2=\M.
\end{equation}

We will use the partial wave decomposition of the amplitudes in the form
\begin{equation}\label{PWD}
A_{i\rightarrow f}(s,\cos\theta)=32\pi\sum_l A_l^{i\rightarrow f}(s) (2l+1)P_l(\cos\theta),
\end{equation}
where $P_l$ are Legendre polynomials.

From the unitarity of S-matrix (and real analyticity of the physical
amplitude as well as $T$-invariance) there follows the unitarity
relations. Assuming that the only relevant intermediate states are those containing two
particles $\alpha$ and $\beta$ with masses $m_\alpha$ and $m_\beta$, its form for the partial wave of
amplitudes reads
\begin{equation}\label{unitarita}
\im A_l^{i\rightarrow
f}(s)=\sum_{\alpha,\beta}\frac{2}{S_F}\,\frac{\lambda^{1/2}_{\alpha\beta}(s)}{s}
A_{l}^{i\rightarrow(\alpha,\beta)}(s)\left[A_{l}^{f\rightarrow(\alpha,\beta)}(s)\right]^*\theta(s-(m_\alpha+m_\beta)^2).
\end{equation}
Here, $S_F$ denotes the symmetry factor ($S_F=1$ if the two intermediate
states are distinguishable and $S_F=2$ if they are not).

Finally, we use the definition of the (signs of) meson fields given in the matrix
\begin{equation}\begin{split}
\phi(x)=\lambda_a \phi_a(x)&=\begin{pmatrix}\pi^0+\frac1{\sqrt3}\eta&-\sqrt 2 \pi^+&-\sqrt 2 K^+\\
\sqrt 2 \pi^-&-\pi^0+\frac1{\sqrt3}\eta&-\sqrt 2 K^0\\
\sqrt 2 K^-&-\sqrt 2 \overline{K^0}&-\frac2{\sqrt3}\eta \end{pmatrix}.
\end{split}\end{equation}
The meson states are chosen so that the matrix element of mesons
gets one minus sign for each charged particle in the final state.

\section{Reconstruction theorem}

\subsection{Statement of the theorem}
In this section we prove the following theorem: The amplitude
$S(s,t;u)$ of a given process $A+B\rightarrow C+D$ fulfilling all
the conditions from the next subsection can be reconstructed to (and excluding)
\Op8 order just from the knowledge of imaginary parts of $s$ and
$p$ partial waves of all crossed amplitudes and some polynomial
\begin{equation}\label{teorem}
\begin{split}
S(s,t;u)&=R_4(s,t;u)+\Phi_0(s)+\left[s(t-u)+\Delt{AB}\Delt{CD}\right]\Phi_1(s)+\Psi_0(t)\\
&\quad+\left[t(s-u)+\Delt{AC}\Delt{BD}\right]\Psi_1(t)
+\Omega_0(u)+\left[u(t-s)+\Delt{AD}\Delt{BC}\right]\Omega_1(u)+\Op8,
\end{split}
\end{equation}
where $R_4(s,t;u)$ is a third-order polynomial in Mandelstam
variables (obeying exactly the same symmetries as the original
amplitude $S(s,t;u)$) and
{\allowdisplaybreaks
\begin{align}
\Phi_0(s)&=32s^3\int_\Sigma^{\Lambda^2}\frac{dx}{x^3}\frac{\im S_0(x)}{x-s}\,,\label{phi0}\\
\Phi_1(s)&=96s^3\int_\Sigma^{\Lambda^2}\frac{dx}{x^3}\frac{1}{x-s}\im\frac{S_1(x)}{\lambda_{AB}^{1/2}(x)\lambda_{CD}^{1/2}(x)}\,,\label{phi1}\\
\Psi_0(t)&=32t^3\int_\tau^{\Lambda^2}\frac{dx}{x^3}\frac{\im T_0(x)}{x-t}\,,\\
\Psi_1(t)&=96t^3\int_\tau^{\Lambda^2}\frac{dx}{x^3}\frac{1}{x-t}\im\frac{T_1(x)}{\lambda_{AC}^{1/2}(x)\lambda_{BD}^{1/2}(x)}\,,\label{psi1}\\
\Omega_0(u)&=32u^3\int_\Upsilon^{\Lambda^2}\frac{dx}{x^3}\frac{\im U_0(x)}{x-u}\,,\label{omega0}\\
\Omega_1(u)&=96u^3\int_\Upsilon^{\Lambda^2}\frac{dx}{x^3}\frac{1}{x-u}\im\frac{U_1(x)}{\lambda_{AD}^{1/2}(x)\lambda_{BC}^{1/2}(x)}\,,\label{omega1}
\end{align}}%
$S_0$ and $S_1$ are $s$ and $p$ partial wave of the process in
question, similarly $T_0$ and $T_1$ and $U_0$ and $U_1$ the partial
waves of the $T$ and $U$ crossed processes. $\Sigma, \tau, \Upsilon$
are the minima of squared invariant mass of all possible
intermediate states in the $S$, $T$ and $U$ channels (see \refer{minimum}).
Finally,  $\Lambda$ is the scale from the assumptions.

\subsection{Assumptions of the theorem}
We use this theorem in the (strong part of) chiral perturbation
theory, but for better understanding of it, we are formulating all
the assumptions in the general way and then discussing their
validity in \chiPT\ separately. The assumptions of the theorem are:
\begin{enumerate}
\item There exists a threshold $\Lambda$ up to which we can regard the
theory under consideration to be complete; this means, for example, that
under this threshold there appears no particle of other types than
those which are already explicitly included in the theory and all the
influence of such extra particles is already taken into account effectively
or negligible under the threshold. In chiral perturbation theory the
threshold $\Lambda$ is the threshold of production of non-pGB
states, i.e.\ $\Lambda\sim1\,\mathrm{GeV}$.

\item There exists a well-behaved expansion of the amplitude in
powers of $p/\Lambda$. In \chiPT\ it is enough to have any
chiral expansion, not necessarily the standard Weinberg one. As is
common in \chiPT, instead of \Opl{n} we write just $\Op{n}$ for
short.

\item We can write a three times\footnote{As discussed later on, the exact number of subtractions is not so relevant.} subtracted dispersion relation for
the amplitude $S(s,t;u)$ in the complex $s$ plane for fixed value of
$u$ (in the form \refer{subtrakce}). This is the essential
assumption of the proof - the theorem stands or falls by it. This assumption is also connected with the validity of the
partial wave expansion (for the sake of simplicity, we assume
that this expansion can be also analytically continued below the
physical threshold where needed). The legitimacy of these
assumptions in the case of \chiPT\ is discussed in
Appendix~\ref{disperzni diskuze}.

\item We further assume that the amplitude considered as a function of a single Mandelstam variable (with the other fixed at some appropriate value) is analytic in some unempty open region. The regions where this and the previous assumptions are valid are discussed (and plotted) in Appendix~\ref{disperzni diskuze}.

\item The crossing relations given in the form of the previous section are
valid, i.e.\ \ $S(s,t,u)=T(t,s,u)=U(u,t,s)$. That is a widely accepted assumption, whose validity for a
general $2\rightarrow2$ process has been proven from axiomatic theory by Bross, Epstein, and Glaser \cite{BEG}.

\item The absorptive parts of the $l\ge2$ partial amplitudes are suppressed to
the \Op8 order - thanks to that we can (up to this order) deal just
with the first two partial waves in the theorem. In our case that
follows from the (pseudo-)Goldstone boson character of the particles
under consideration.
In \chiPT\ the amplitudes behave dominantly as \Op2 and do not
contain any bound state poles. The unitarity relation
\refer{unitarita} then implies that its imaginary part behaves as
\Op4, i.e.\ the amplitudes are dominantly real. According to the
analyticity of the amplitude, its leading \Op2 part should be a
polynomial in Mandelstam variables. Moreover, it has to be a
polynomial of the first order; otherwise its coefficient would grow
up as mass of the pGB went to zero, and that would contradict the
finiteness of the S-matrix in the chiral limit with the external
momenta fixed\footnote{We will see that the finiteness of the
S-matrix in the chiral limit deserves to be regarded as one of the
full-valued assumptions by itself.}. However, the first-order
polynomial could not contribute to $l\ge2$ partial waves, thus these
partial waves should behave in the chiral limit as \Op4, thereby
again using the unitarity relation, the imaginary parts of these partial
waves are at least of the \Op8 order.
\end{enumerate}

\subsection{Proof of the theorem}
One of our assumptions is that for the amplitude we can write the
three times subtracted dispersion relation in the form:
\begin{multline}\label{subtrakce}
S(s,t;u)=P_3(s,t;u)+\frac{s^3}{\pi}\int_{\Sigma}^{\infty}\frac{dx}{x^3}\frac{\im S(x,\M-x-u;u)}{x-s}\\
+\frac{t^3}{\pi}\int_{\tau}^{\infty}\frac{dx}{x^3}\frac{\im S(\M-x-u,x;u)}{x-t}\,,
\end{multline}
where $P_n(s,t;u)$ is a polynomial of the $(n-1)$-th order in $s$
and $t$.\footnote{The coefficients of this polynomial depend on $u$. Thanks to the relation for the sum of the
Mandelstam variables, such polynomials could be written in the form
\begin{equation}\label{rozvoj polynomu}
P_n(s,t;u)=\alpha(u)+\beta(u)(s-t)+\gamma(u)(s-t)^2+\dots+\omega(u)(s-t)^{n-1}.
\end{equation}}
$\Sigma$ is the minimum of the squared invariant mass of all the
possible intermediate states $(\alpha,\beta)$ in the S
channel (or rather the lowest mass
of the state with the same quantum numbers as the \emph{in} and
\emph{out} state in the S-channel)\footnote{In this relation we anticipate again that we will
consider the problems under conditions implying relevance of the
two-particle intermediate states only.},
\begin{equation}\label{minimum}
\Sigma=\min_{(\alpha,\beta)}(m_\alpha+m_\beta)^2,
\end{equation}
and analogically $\tau$ for the $T$ channel (later on we will also need
the same for the $U$ channel, what will be denoted by $\Upsilon$).

Thanks to the crossing symmetry we can replace the $S(s,t;u)$
amplitude in the second integral with the $T(t,s;u)$. Further, since
we assume that we know the complete theory only up to some threshold
$\Lambda$, we have to split the dispersion integral into two parts -
the low-energy $(x\le\Lambda^2)$ and the high-energy
$(x\ge\Lambda^2)$ part. We consider the region where
$s\ll\Lambda^2$, i.e.\ in the high-energy part, $s\ll x$, we
have\footnote{The fact that the remainder $\sum_{n\ge4}s^n R_n(u)$
is of order at least \Op8 follows again from the finiteness of the
amplitude in the chiral limit (the importance of this assumption
stems from here) and the crossing symmetry.}
\begin{equation}\label{odrezani}
\frac{s^3}{\pi}\int_{\Lambda^2}^{\infty}\frac{dx}{x^3}\left.\frac{\im
S(x,\M-x-u;u)}{x-s}\right|_{s<<x}=s^3 H_\Lambda(u)+\Op8,
\end{equation}
where
\begin{equation}
H_\Lambda(u)=\frac1{\pi}\int_{\Lambda^2}^{\infty}\frac{dx}{x^3}\frac{\im S(x,\M-x-u;u)}{x}
\end{equation}
is a function of $u$ only. We can proceed similarly for the high-energy part of
the integration of $T(s,t;u)$. Hence, these high-energy parts can be
to the order \Op8 added to the subtraction polynomial now thereby
of the third order\footnote{The consequence of that is also that we
can formally extend the validity of the original theory even beyond
threshold $\Lambda$ and compute the dispersion integral up to
infinity, pretending that the original theory is really complete
(naturally with the appropriate modification of the third-order
polynomial). Similar arguments would appear even if we introduced
the cutoff $\Lambda$ by other reasons, e.g.\ by reason of numerical
integration.}. Afterwards, the dispersion relation is of the form
\begin{multline}\label{imaginary PWD}
S(s,t;u)=P_4(s,t;u)+\frac{s^3}{\pi}\int_{\Sigma}^{\Lambda^2}\frac{dx}{x^3}\frac{\im S(x,\M-x-u;u)}{x-s}\\
+\frac{t^3}{\pi}\int_{\tau}^{\Lambda^2}\frac{dx}{x^3}\frac{\im T(x,\M-x-u;u)}{x-t}+\Op8.
\end{multline}

We decompose the (imaginary parts of) amplitudes into the partial
waves
\begin{equation}\label{decomposition of PW}
\begin{split}
S(s,t;u)&=32\pi\left(S_0(s)+3\cos\theta S_1(s)+S_{l\ge2}(s,t;u)\right)\\
&=32\pi\left(S_0(s)+3\frac{s(t-u)+\Delt{AB}\Delt{CD}}{\lambda_{AB}^{1/2}(s)\lambda_{CD}^{1/2}(s)}S_1(s)+S_{l\ge2}(s,t;u)\right).
\end{split}
\end{equation}
The terms $S_{l\ge2}(s,t;u)$ incorporate contributions of all the
higher $(l\ge2)$ partial waves and are suppressed to \Op8 according
to our assumptions.

From the integrals of the P partial waves, we can extract functions
depending just on $u$ and include them in the polynomial, e.g.\
\begin{multline}
\int_{\Sigma}^{\Lambda^2}\frac{dx}{x^3}\frac{x(s+t-x-u)}{x-s}\im\frac{S_1(x)}{\lambda_{AB}^{1/2}(x)\lambda_{CD}^{1/2}(x)}\\
=\int_{\Sigma}^{\Lambda^2}\frac{dx}{x^3}\frac{x(t-u)}{x-s}\im\frac{S_1(x)}{\lambda_{AB}^{1/2}(x)\lambda_{CD}^{1/2}(x)}-\int_{\Sigma}^{\Lambda^2}\frac{dx}{x^2}\im\frac{S_1(x)}{\lambda_{AB}^{1/2}(x)\lambda_{CD}^{1/2}(x)}.
\end{multline}

Finally, we use this rearrangement of the fractions
\begin{equation}\label{rearrangement}
\frac{x(t-u)}{x-s}=\frac{s(t-u)}{x-s}+t-u,
\end{equation}
where the second term gives after integration a contribution in the
form $t-u$ times a function of $u$ only and therefore as a
polynomial in $t$ can also be included into the subtraction
polynomial $P_4$.

After that, the amplitude reads
\begin{multline}
S(s,t;u)=P_4(s,t;u)+\Phi_0(s)+[s(t-u)+\Delt{AB}\Delt{CD}]\Phi_1(s)\\
+\Psi_0(t)+[t(s-u)+\Delt{AC}\Delt{BD}]\Psi_1(t)+\Op8,
\end{multline}
where $\Phi_0(s),\Phi_1(s),\Psi_0(t)$, and $\Psi_1(t)$ are given by \refer{phi0} -- \refer{psi1}.

In order to implement the $s\leftrightarrow u$ crossing symmetry, we
add and subtract the following polynomial in $s$ and $t$ (with
coefficients depending on $u$)
\begin{equation}
\Omega_0(u)+[u(t-s)+\Delt{AD}\Delt{BC}]\Omega_1(u),
\end{equation}
with $\Omega_0(u)$ and $\Omega_1(u)$ given by \refer{omega0} and
\refer{omega1} and we include the subtraction of this polynomial in
the polynomial $P_4(s,t;u)$ and thereby get the amplitude in the
symmetric form \refer{teorem}.

Until now, we have proved the whole theorem except for the
$u$-dependence of the polynomial $P_4(s,t;u)$. However, we could write similar
relations also for the amplitudes $T(s,t;u)$ and $U(s,t;u)$ with the
polynomials in $(s,u)$ and the coefficients depending on variable
$t$. Thanks to the crossing symmetry and to the symmetric form of
the rest of the formula, these polynomials should obey the same
crossing symmetry. From this and from the analyticity of the
amplitudes in a neighborhood of $u$ for some fixed $s$ in our
kinematic region, we can conclude that $P_4(s,t;u)$ are
polynomials in respect to all Mandelstam variables.

\section{Application of the theorem to the 4-meson processes to the \Op4 order}
As we have already outlined, in the case of $2\rightarrow2$
scattering processes of pseudoscalar octet mesons, we can use the
theorem and the unitarity relation to a selfconsistent construction
of all those amplitudes to (and excluding) \Op8 order. We need just a
parametrisation of \Op2 amplitudes as its input.

In the following we will consider the strong isospin conservation
limit\footnote{Recall that in this case the validity of the required
dispersion relations can be proven directly from the axiomatic theory as discussed in Appendix~\ref{disperzni diskuze}.}, where
Ward identities imply that there are just 7 independent processes.
From them the amplitudes of all the possible physical processes can
be determined as summarised in Appendix~A. Since we want to keep the
independence on the specific power-counting scheme, we have
parameterised the \Op2 amplitudes in the most general form obeying
their symmetries (crossing, Bose, and isospin) - given also in Appendix~A. Inserting them into the
unitarity relations \refer{unitarita} (assuming that we have
amplitudes of all the needed intermediate processes), we get the
imaginary parts to \Op4 order, which can be installed into the
theorem and we receive thereby the amplitudes in that order. With
the same procedure (though more technically involved), we can proceed
with the second iteration and obtain thereby the amplitudes to (and excluding)
\Op8 order. For the sake of simplicity, in this paper we confine
ourselves to the first iteration and thus to the results to \Op4
order.

It remains to show that we already have the amplitudes of all the
relevant intermediate processes, i.e.\ the self-consistency of
this procedure. We are concerned with the processes to (and excluding) \Op8
chiral order and (in accordance with the first assumption of the
theorem) in the region bellow the appearance of all non-Goldstone
particles and thereby away from poles and cuts of such intermediate
states. Therefore, their effect is included just in the polynomial
of the reconstruction theorem and in the higher orders. The
intermediate states with an odd number of Goldstone bosons are
forbidden by the even intrinsic parity of the effective Lagrangian -
because the effect of the axial anomaly enters at the \Op4 order
(Wess, Zumino and Witten \cite{WZW}), such intermediate states
induce the contribution of the order at least \Op8. Furthermore, the contribution
of the states with more than two Goldstone bosons is also suppressed
to \Op8 order since the $n$-Goldstone-boson invariant phase space
scales like $p^{2n-4}$ and amplitudes with an arbitrary number of
external Goldstone boson legs behave dominantly as \Op2.
Consequently, the contribution of such ($n>2$) intermediate states
to the imaginary part of the amplitude is according to the unitarity
relation at least \Op8. Thus, we can consider the
two-Goldstone-boson intermediate states only.

Further simplification in the \Op4 order computation appears. Using
again the arguments of finiteness of S-matrix in the chiral limit
with the external momenta fixed, the coefficients of the polynomial
in the theorem should be at least \Op0, and so the polynomial in the
\Op4 amplitude is maximally of the second order in Mandelstam
variables.

\subsection{The application schematically}
The \Op2 amplitudes can be written (cf.\ with \refer{rozvoj polynomu}, and Appendix~A with use of \refer{soucet Mandelstamu}) in the form\footnote{Let us remind that in the \Op2 case $\alpha(s)=\alpha_0+\alpha_1 s$, i.e.\ it is just the first order polynomial in $s$.}
\begin{equation}
A=\frac{1}{F_\pi^2}\left(\alpha(s)+\beta(t-u)\right).
\end{equation}
The decomposition into the partial waves can be obtained using \refer{uhel} and \refer{decomposition of PW}
\begin{align}
A_0&=\frac{1}{32\pi}\frac{1}{F_\pi^2}\left(\alpha(s)-\beta\frac{\Delt{AB}\Delt{CD}}{s}\right),\\
A_1&=\frac{1}{32\pi}\frac{1}{3F_\pi^2}\beta\frac{\lambda^{1/2}_{AB}(s)\lambda^{1/2}_{CD}(s)}{s}\,.
\end{align}
The (right) discontinuity of our function $S_0$ and $S_1$ of the theorem, given by the unitarity relation,
is therefore in the $\Op4$ case very simple:
\begin{align}
\begin{split}
\im S_0(s)&=\sum_{\substack{\scriptscriptstyle k-\text{interm.}\\\scriptscriptstyle\text{states}}}
\frac{2}{S_{F_k}}\frac{\lambda_k^{1/2}(s)}{s}\left(\frac{1}{32\pi}\right)^2\frac{1}{F_\pi^4}\left[\left(\alpha_{ik}(s)-\beta_{ik}\frac{\Delt{AB}\Delta_{k}}{s}\right) \left(\alpha_{kf}(s)-\beta_{kf}\frac{\Delta_{k}\Delt{CD}}{s}\right)\right]\\
&\qquad\qquad\qquad\times\theta(s-s^t_k),
\end{split}\\
\im S_1(s)&=\sum_{\substack{\scriptscriptstyle k-\text{interm.}\\\scriptscriptstyle\text{states}}}
\frac{2}{S_{F_k}}\frac{\lambda_k^{1/2}(s)}{s}\left(\frac{1}{32\pi}\right)^2\frac{1}{9F_\pi^4}\bigg[\beta_{ik}\frac{\lambda^{1/2}_{AB}(s)\lambda^{1/2}_{k}(s)}{s}\beta_{kf}\frac{\lambda^{1/2}_{k}(s)\lambda^{1/2}_{CD}(s)}{s}\bigg]\theta(s-s^t_k),
\end{align}
where the sums go over all the possible (two-pGB) intermediate states $k$ with its symmetry factor $S_{F_k}$ and all the objects with lower index $k$ are the quantities relating to this intermediate state, similarly indices $i$ and $f$ refer to initial and final state. Finally $s^t_k$ is a threshold of this intermediate state.

Thanks to the behaviour of square root of the triangle function, $\lambda^{1/2}(s)\rightarrow s$ for $s\rightarrow\infty$ and the fact that $\alpha(s)$ is the polynomial maximally of the first order in $s$, we know that not all the terms (in the brackets) need all the subtractions to give finite integrals (\ref{phi0}) -- (\ref{omega1}) (e.g.\ the parts with negative power of $s$ in the polynomial in the brackets do not even need any subtraction). Any subtraction redundant in this aspect then gives just a polynomial, which can be included into the polynomial $R_4(s,t,t)$ of the theorem (\ref{teorem}) and into the contributions of higher orders. Furthermore, if the integrals rising on both sides of rearrangement (\ref{rearrangement}) have a good mathematical sense, we can pull out the polynomial in front of the integral again with possible change of the polynomial $R_4(s,t,t)$. Therefore, applying the reconstruction theorem, we can write $\Phi_0$ and $\Phi_1$ in the minimal\footnote{With the minimal number of subtraction needed to get a finite nonpolynomial part, i.e.\ here with one or two subtractions.} form (except for the polynomial included to $R_4(s,t,t)$):
\begin{multline}
\Phi_0(s)=\frac{1}{16\pi^2}\frac{1}{F_\pi^4}\sum_{k}\frac{1}{S_{F_k}}\Bigg(\alpha_{ik}(s)\alpha_{kf}(s)s\int_{s^t_k}^{\infty}\frac{dx}{x}\frac{1}{x-s}\frac{\lambda_k^{1/2}(x)}{x}\\
-\big(\alpha_{kf}(s)\beta_{ik}\Delt{AB}+\alpha_{ik}(s)\beta_{kf}\Delt{CD}\big)\Delta_{k}\int_{s^t_k}^{\infty}\frac{dx}{x}\frac{1}{x-s}\frac{\lambda_k^{1/2}(x)}{x}\\
+\beta_{ik}\beta_{kf}\Delta_{k}^2\Delt{AB}\Delt{CD}\int_{s_k^t}^{\infty}\frac{dx}{x^2}\frac{1}{x-s}\frac{\lambda_k^{1/2}(x)}{x}\Bigg).
\end{multline}

Similarly for the P-wave function $\Phi_1$ (in addition we have used $\lambda_k(x)=x^2-2x\Sigma_k+\Delta_k^2$, where $\Sigma_k$ is the sum of the second powers of masses of the two particles in the intermediate state $k$):
\begin{equation}
\Phi_1(s)=\frac{1}{16\pi^2}\frac{1}{3F_\pi^4}\sum_{k}\frac{1}{S_{F_k}}\beta_{ik}\beta_{kf}\Bigg(\big(s-2\Sigma_k\big) \int_{s^t_k}^{\infty}\frac{dx}{x}\frac{1}{x-s}\frac{\lambda_k^{1/2}(x)}{x}+\Delta_k^2\int_{s^t_k}^{\infty}\frac{dx}{x^2}\frac{1}{x-s}\frac{\lambda_k^{1/2}(x)}{x}\Bigg).
\end{equation}

In the integrals we recognize the once and twice subtracted dispersion integrals from Appendix~B (\refer{once subtracted} and \refer{twice subtracted}), i.e.\
\begin{align}
\begin{split}
\Phi_0(s)&=\frac{1}{F_\pi^4}\sum_{k}\frac{1}{S_{F_k}}\Bigg(\beta_{ik}\beta_{kf}\frac{\Delta_{k}^2}{s^2}\Delt{AB}\Delt{CD}\overline{\overline{J}}_k(s)\\
&\qquad+\bigg(\alpha_{ik}(s)\alpha_{kf}(s)-(\alpha_{kf}(s)\beta_{ik}\Delt{AB}+\alpha_{ik}(s)\beta_{kf}\Delt{CD})\frac{\Delt{k}}{s}\bigg)\overline{J}_k(s)\Bigg),
\end{split}\\
\Phi_1(s)&=\frac{1}{3F_\pi^4}\sum_{k}\frac{1}{S_{F_k}}\beta_{ik}\beta_{kf}\bigg(\bigg(1-2\frac{\Sigma_k}{s}\bigg)\overline{J}_k(s)+\frac{\Delta_k^2}{s^2}\overline{\overline{J}}_k(s)\bigg).
\end{align}

Using the change of the polynomial for the last time (in the replacement of $\frac{\overline{\overline{J}}(s)}{s}$ with $\frac{\overline{J}(s)}{s}$), we have the result:
\begin{equation}
\begin{split}
&\Phi_0(s)+[s(t-u)+\Delt{AB}\Delt{CD}]\Phi_1(s)\\
&\quad=\frac{1}{3F_\pi^4}\sum_{k}\frac{1}{S_{F_k}}\Bigg(4\Delt{AB}\Delt{CD} \beta_{ik}\beta_{kf}\Delta_{k}^2\frac{\overline{\overline{J}}_k(s)}{s^2}\\
&\qquad+\bigg(\beta_{ik}\beta_{kf}\big((t-u)\Delta_k^2-2\Delt{AB}\Delt{CD}\Sigma_k\big) -3(\alpha_{kf}(s)\beta_{ik}\Delt{AB}+\alpha_{ik}(s)\beta_{kf}\Delt{CD})\Delta_{k}\bigg) \frac{\overline{J}_k(s)}{s}\\
&\qquad+\bigg(3\alpha_{ik}(s)\alpha_{kf}(s)+\beta_{ik}\beta_{kf}\big((t-u)(s-2\Sigma_k)+\Delt{AB}\Delt{CD}\big)\bigg) \overline{J}_k(s)\Bigg).
\end{split}
\end{equation}

We get similar results also for the $T$- and $U$-crossed parts\footnote{If we carry out the same extraction of polynomial also for them, we get again these results by the change $s\leftrightarrow t$ respectively $s\leftrightarrow u$ and appropriate permutation in ABCD and then the symmetries of the polynomial of theorem remain the same.}. Final results of this procedure for all the amplitudes are given in Appendix~B.

\section{Summary and discussions}
In this paper we have worked out a generalisation of the dispersive
analysis from \cite{SSF} to the processes involving all the light
octet mesons. We have formulated the reconstruction theorem for
their amplitudes free of the particular isospin structure of the
processes (which is used afterwards for the individual processes
separately) and have provided (an outline of) its proof. We have
also attempted to point out exactly the properties of the theory
which are needed for its proof mainly due to better understanding of
it as well as of its further generalisation on more complicated
cases (as below). In our particular case we have succeeded in
justifying most of them practically from the first principles of
axiomatic quantum field theory - at least in the regions from Fig.~1.

The amplitudes of 4-octet-meson processes (on condition of the
strong isospin conservation) form a selfconsistent system with
respect to the theorem and the unitarity relations, thanks to the fact that
we can simply construct these amplitudes to (and excluding\footnote{Since we also want to include the power countings where the contributions of odd chiral orders appear (like generalised chiral power counting), we write instead of valid to \Op6 order (what will be case if odd orders do not appear) or to \Op7 order (where the odd orders are included) just valid to and excluding \Op8 order to avoid misunderstanding. We should also remind that in these odd chiral order power countings, the one-loop result is not \Op4 but \Op5 -- we can get such a result by simple modification of the results presented keeping the $\beta_{\pi\eta}$ and $\gamma_{KK}$ parameters also in the intermediate processes.}) \Op8 order (in this
work explicitly computed only to \Op4 order) independently on the
specific power-counting model in use. The amplitudes for the
particular model can be obtained with the specific choice of the parameters. For example, we can compare the nonpolynomial part of our results (with the standard choice of \Op2 constants from Appendix~A) with that part of the standard chiral perturbation results (given by \cite{Spanele}) -- we see an agreement except for the different sign convention\footnote{When
comparing, one should pay attention to their caption of the 4-kaon
process $\Ka K^0\rightarrow K^+ K^-$, which gives us the false
illusion about the value of the Mandelstam variables, whereas the
right value would be better evoked by the caption $\Ka
K^0\rightarrow K^- K^+$.}. We can also use their result to get a relation between our \Op4 polynomial parameters and low-energy constants (LECs)
of the standard chiral perturbation theory in the particular renormalisation scheme used therein just by comparison of the polynomial parts, as is discussed and given in Appendix~C.\footnote{If we used the \Op4 parameters that has been obtained thereby also in the unitarity part, we would be already counting a part of the \Op6 corrections.} It is worth mentioning that by subtracting the center of the Mandelstam triangle in the second-order polynomial part of our results, the parameters of the first-order polynomial, i.e.\ $\alpha$'s, $\beta$'s and $\gamma$'s, do not depend in the subtraction scheme used by \cite{Spanele} in \Op4 order on the LECs of operators with four derivatives from $\mathcal{L}_4$ in \cite{Gasser}, i.e.\ $L_1^r$, $L_2^r$, and $L_3$.

Some of the computed amplitudes ($KK\rightarrow$ anything, processes
involving $\eta$'s) have not been computed in such a general form
yet. However, this pioneerness is connected with one problem of the
assumptions of the theorem. We have assumed the existence of a
threshold separating the processes containing just those 8 mesons
from the processes in which particles of other types (e.g.\
resonances) appear, and then we focus on a kinematical range far
below that threshold. However, in the case of the processes with the
exception of $\pi\pi\rightarrow\pi\pi$ and $\pi K\rightarrow\pi K$,
this range is very narrow or does not even exist (e.g.\ in the
process $KK\rightarrow KK$ with a kinematical threshold of
$990\,\mathrm{MeV}$ there can appear the $\rho$ resonance with its
mass of $770\,\mathrm{MeV}$). The significance of our result thereby
decreases. Nevertheless, beside the elegance of our construction due
to its intrinsic self-consistency, our results can become useful for
a check of complicated models or simulations, where one can separate
(or completely turn off) the effects of these resonances. (The
advantage is that even in the most general case in (to and excluding) \Op8 order, there
is a small number ($=47$) of parameters, coupling different processes
together.)

To that end it would be naturally preferable to perform also the
second iteration and get thereby the whole information achievable
from the reconstruction theorem. As has been already mentioned, the
completion of the second iteration is conceptually straightforward; however, there appear certain technical complications which are connected with the computation of partial waves of
amplitudes, where we need to compute integrals of the type $\int t^n
\overline{J}(t)dt$ and then to \lq\lq dispersively integrate" them. Because of the appearance of the three different masses, the results are much more complicated than in the 2 flavour $\pi\pi$ case (with only one mass) and so will be presented with all the technical details in a separate paper.

The careful reader has surely noticed that in the relation
\refer{imaginary PWD} or \refer{phi1} and the other similar (in
opposition to the verbal formulation of the theorem), we have
emphasised that the required imaginary part is not of the partial
wave but of the partial wave divided by
$\lambda_{AB}^{1/2}\lambda_{CD}^{1/2}$, i.e.\ in principle in the higher order there
could appear a problem under the physical threshold of the
amplitudes where $\lambda_{AB}^{1/2}\lambda_{CD}^{1/2}$ could be
imaginary. There presents itself a question whether we can obtain
this quantity also from the unitarity relations. But we should
remind that the origin of the relation \refer{unitarita} where the
imaginary part of the partial wave times the relevant Legendre
polynomial also occurs. So the only question is whether we can
believe the analytical continuation of the partial wave
decomposition and of the unitarity relation under the physical
threshold. If we do so, then the unitarity relation gives exactly
what we need to the theorem, i.e.\ what we call the imaginary part of
partial wave for simplicity.

The further important question is of the number of subtractions in
the dispersive relations. Jin and Martin \cite{Jin Martin} have shown
that thanks to the Froissart bound \cite{Froissart} it suffices to
consider just two subtractions. However, the Froissart bound deals
with the complete theory (full QCD), and in an effective theory one
does not know (or more exactly one does not deal with) what is above
$\Lambda$, and so the situation can occur that more than two subtractions are needed. In other words Froissart tells us two subtractions are
sufficient if we supply the dispersive integral above $\Lambda$ with
the complete theory. How to deal with this unknown, we have
already seen near the relation \refer{odrezani} -- for $n$
subtractions we get a series of infinite order beginning with
the $n$-th power of Mandelstam variable. Fortunately, we have shown
that the terms with more than fourth power are at least of \Op8
order. And so, if we assume the finiteness of the S-matrix in the
chiral limit (and the other assumptions used there), we end up with
the third-order polynomial and the remainder of \Op8 order no matter
with how many subtraction we have begun.

From the possible extensions of the theorem we find especially
interesting two of them, the extension to the analysis of the meson
formfactors and the computation of amplitudes of $K\rightarrow3\pi$
decays (with appearance of cusps). The latter of them is the subject
we are recently working on \cite{nase proceedings,nas cusp}.

\section*{Acknowledgment}
We are grateful to K.\ Kampf for careful reading of the manuscript and helpful suggestions. We also thank to M.\ Knecht for inspiring discussions.

The work was supported by the Center for Particle Physics, Project
No.\ LC527 of the Ministry of Education of the Czech Republic and by the EU-RTN Programme, Contract No.\ MRTN--CT-2006-035482, \lq\lq Flavianet."

M.Z.\ would like also gratefully acknowledge an Early Stage
Researcher position supported by the EU-RTN Programme, Contract No.\
MRTN--CT-2006-035482, \lq\lq Flavianet."

\appendix
\section{\Op2 amplitudes and the symmetry properties of amplitudes of the physical mesons}\label{op2}
From the isospin Ward identities it follows that all the physical amplitudes could be expressed in terms of the 7 independent amplitudes. In our convention:
\begin{enumerate}
\item $\eta\eta\rightarrow\eta\eta$
\begin{equation}
A(\eta\eta\rightarrow\eta\eta)=A_{\eta\eta}(s,t;u).
\end{equation}

\item $\pi\eta\rightarrow\pi\eta$
\begin{align}
A(\pi^\pm\eta\rightarrow\pi^\pm\eta)&=A_{\pi\eta}(s, t;u),\\
A(\pi^0\eta\rightarrow\pi^0\eta)&=A_{\pi\eta}(s,t;u).
\end{align}

\item $\pi\pi\rightarrow\pi\pi$
\begin{align}
A(\pi^+\pi^-\rightarrow\pi^0\pi^0)&=-A_{\pi\pi}(s, t;u),\\
A(\pi^+\pi^-\rightarrow\pi^+\pi^-)&=A_{\pi\pi}(s, t;u)+A_{\pi\pi}(t,s;u),\\
A(\pi^0\pi^0\rightarrow\pi^0\pi^0)&=A_{\pi\pi}(s,t;u)+A_{\pi\pi}(t,s;u)+A_{\pi\pi}(u,t;s).
\end{align}

\item $KK\rightarrow\eta\eta$
\begin{equation}
A(\Ka K^0\rightarrow\eta\eta)=-A(K^- K^+\rightarrow\eta\eta)=A_{\eta K}(s,t;u).
\end{equation}

\item $KK\rightarrow\pi\eta$
\begin{align}
A(K^-K^+\rightarrow\pi^0\eta)&=-A_{\eta\pi K}(s,t;u),\\
A(\Ka K^0\rightarrow\pi^0\eta)&=-A_{\eta\pi K}(s,t;u),\\
A(K^-K^0\rightarrow\pi^-\eta)&=-\sqrt 2A_{\eta\pi K}(s,t;u),\\
A(\Ka K^+\rightarrow\pi^+\eta)&=-\sqrt 2A_{\eta\pi K}(s,t;u).
\end{align}

\item $\pi\pi\rightarrow KK$
\begin{align}
A(\pi^0\pi^0\rightarrow \Ka K^0)&=-A(\pi^0\pi^0\rightarrow K^- K^+)=A_{\pi K}(s,t;u),\\
A(\pi^-\pi^0\rightarrow K^- K^0)&=\sqrt 2B_{\pi K}(s,t;u),\\
A(\pi^+\pi^0\rightarrow \Ka K^+)&=-\sqrt 2B_{\pi K}(s,t;u),\\
\label{pipiKK souctova amplituda}
A(\pi^-\pi^+\rightarrow K^- K^+)&=A_{\pi K}(s,t;u)+B_{\pi K}(s,t;u),\\
A(\pi^-\pi^+\rightarrow \Ka K^0)&=-A_{\pi K}(s,t;u)+B_{\pi K}(s,t;u).
\end{align}
For these processes there are two important amplitudes -- the symmetric one $A_{\pi K}(s,t,u)$ and the antisymmetric one $B_{\pi K}(s,t,u)$. In fact, we can consider the only independent amplitude as \refer{pipiKK souctova amplituda} and then extract these amplitudes as the symmetric and the antisymmetric part of it in respect to the exchange of $t$ and $u$.

\item $K K\rightarrow KK$
\begin{align}
A(K^-K^+\rightarrow \Ka K^0)&=-A_{KK}(s,t;u),\label{lisici se K}\\
A(K^-K^+\rightarrow K^- K^+)&=A(\Ka K^0\rightarrow \Ka K^0)=A_{KK}(s,t;u) +A_{KK}(t,s;u).\label{stejna K}
\end{align}
\end{enumerate}

\subsection*{The most general form of the \Op2 (\Op3) amplitudes}
The \Op2 amplitudes could be constructed as the most general
invariant amplitudes satisfying the crossing, Bose and isospin symmetries and using the
fact that they should be polynomials of the first order in the
Mandelstam variables obeying (\ref{soucet Mandelstamu})\footnote{In fact, it is the form of the \Op3
amplitudes, because this should hold also for them.}. This form is
independent of the power counting used:

\begin{equation}
A_{\eta\eta}=-\frac{M_\pi^2}{3F_\pi^2}\alpha_{\eta\eta}\left(1-\frac{4M_\eta^2}{M_\pi^2}\right),
\end{equation}
\begin{equation}
A_{\pi\eta}=\frac{1}{3F_\pi^2}\left(\beta_{\pi\eta}(3t-2M_\eta^2-2M_\pi^2)+\alpha_{\pi\eta}M_\pi^2\right),
\end{equation}
\begin{equation}
A_{\pi\pi}=\frac1{3F_\pi^2}\left(\beta_{\pi\pi}\left(3s-4M_\pi^2\right)+\alpha_{\pi\pi}M_\pi^2\right),
\end{equation}
\begin{equation}
A_{\eta K}=\frac1{4F_\pi^2}\left[\beta_{\eta K}(3s-2M_K^2-2M_\eta^2)+\alpha_{\eta K}\left(2M_\eta^2-\frac 23M_K^2\right)\right],
\end{equation}
\begin{equation}
A_{\eta\pi K}=\frac{1}{4\sqrt{3}F_\pi^2}\left[\beta_{\eta\pi K}\left(3s-2M_K^2-M_\pi ^2-M_\eta^2\right)-\left( 2M_K^2-M_\pi^2-M_\eta^2+\alpha_{\eta\pi K} M_\pi^2\right)\right],
\end{equation}
\begin{equation}
A_{\pi K}=\frac1{12F_\pi^2}\left[\beta_{\pi K}\left(3s-2M_K^2-2M_\pi^2\right)+2(M_K-M_\pi)^2+4\alpha_{\pi K}M_\pi M_K\right],
\end{equation}
\begin{equation}
B_{\pi K}=\frac{1}{4F_\pi^2}\gamma_{\pi K}(t-u),
\end{equation}
\begin{equation}
A_{KK}=\frac1{6F_\pi ^2}\left(\beta_{KK}\left(4M_K^2-3u\right)+3\gamma_{KK}(s-t)+2\alpha _{KK}M_K^2\right).
\end{equation}
After a deeper analysis one can show that in $\Op2$ order the amplitude
$A_{\pi\eta}$ is $t$-independent, so the constant $\beta_{\pi\eta}$
is of the $\Op3$ order and similarly for $\gamma_{KK}$, which is
also equal to zero in the \Op2 case in an arbitrary power counting.
So, generally
\begin{equation}
\beta_{\pi\eta}\sim\Op1\qquad\text{and}\qquad \gamma_{KK}\sim\Op1
\end{equation}

The particular power countings differ in the particular values of
the (13+2) constants in those relations. For example, the standard
power counting gives (to \Op3):

\begin{align}
\alpha_{\eta\eta}^{st}=\alpha_{\pi\eta}^{st}=\alpha_{\pi\pi}^{st}=\alpha _{\eta K}^{st}=
\alpha_{\pi K}^{st}=\alpha_{KK}^{st}=1,\\
\beta_{\pi\pi}^{st}=\beta _{\eta K}^{st}=\beta_{\eta\pi K}^{st}=\beta_{\pi K}^{st}=\gamma_{\pi K}^{st}=\beta_{KK}^{st}=1,\\
\alpha_{\eta\pi K}^{st}=0,\\
\beta_{\pi\eta}^{st}=\gamma_{KK}^{st}=0.
\end{align}

\section{Results}\label{results}
In the results we have denoted the once subtracted integral
\begin{equation}\label{once subtracted}
\overline{J}_{PQ}(s)=\frac{s}{16\pi^2}\int_{\Sigma}^\infty \frac{dx}{x}\frac{1}{x-s}\frac{\lambda^{1/2}_{PQ}(x)}{x}
\end{equation}
in the S channel (analogically for the other crossed channels) and similarly the twice subtracted integral
\begin{equation}\label{twice subtracted}
\overline{\overline{J}}_{PQ}(s)=\frac{s^2}{16\pi^2}\int_{\Sigma}^\infty \frac{dx}{x^2}\frac{1}{x-s}\frac{\lambda^{1/2}_{PQ}(x)}{x}.
\end{equation}

The amplitudes using the first iteration of the reconstruction
procedure read:

$\bullet\quad\eta\eta\rightarrow\eta\eta$
\begin{equation}\begin{split}
A_{\eta\eta}&=\frac{1}{3F_\pi^2}\alpha_{\eta\eta}\left(4M_\eta^2-M_\pi^2\right)+\frac{1}{3F_\pi^4}\delta_{\eta\eta}(s^2+t^2+u^2)\\
&\quad+\left[\frac{1}{6F_\pi^4}\left(Z_{\eta\eta}^{\eta\eta}\overline{J}_{\eta\eta}(s)+Z^{\pi\pi}_{\eta\eta}\overline{J}_{\pi\pi}(s)+Z^{KK}_{\eta\eta}(s)\overline{J}_{KK}(s)\right)\right]+[s\leftrightarrow t]+[s\leftrightarrow u]+\Op5.
\end{split}\end{equation}
\begin{align}
Z_{\eta\eta}^{\eta\eta}&=\frac{1}{3}\alpha_{\eta\eta}^2\left(M_\pi^2-4M_\eta^2\right)^2,\\
Z^{\pi\pi}_{\eta\eta}&=\alpha_{\pi\eta}^2M_\pi^4,\\
Z^{KK}_{\eta\eta}(s)&=\frac{3}{4}\left(3\beta_{\eta K}s-2\left(\beta_{\eta K}+\frac13\alpha_{\eta K} \right) M_K^2+2\left(\alpha_{\eta K}-\beta_{\eta K}\right)M_\eta^2 \right)^2.
\end{align}

\

$\bullet\quad\pi^0\eta\rightarrow\pi^0\eta$
\begin{equation}\begin{split}
A_{\pi\eta}&=\frac{1}{3F_\pi^2}\Big(\beta_{\pi\eta}(3t-2M_\eta^2-2M_\pi^2)+M_\pi^2\alpha_{\pi\eta}\Big)+\frac{1}{3F_\pi^4}\left(\delta_{\pi\eta}(s^2+u^2)+\varepsilon_{\pi\eta} t^2\right)\\
&\quad+\frac{1}{72F_\pi^4}\Big(Z^{KK}_{\pi\eta}(t)\overline{J}_{KK}(t)+Z^{\eta\eta}_{\pi\eta}\overline{J}_{\eta\eta}(t)
+Z^{\pi\pi}_{\pi\eta}(t)\overline{J}_{\pi\pi}(t)\Big)\\
&\quad+\left[\frac1{9F_\pi^4}\left(
Y_{\pi\eta}^{\pi\eta}\overline{J}_{\pi\eta}(s)+Y^{KK}_{\pi\eta}(s)\overline{J}_{KK}(s)\right)\right]+[s\leftrightarrow u]+\Op5.
\end{split}\end{equation}
\begin{align}
\begin{split}
Z^{KK}_{\pi\eta}(t)&=\left(9\beta_{\eta K}t-2(3\beta_{\eta K}+\alpha_{\eta K})M_K^2+6(\alpha_{\eta K}-\beta_{\eta K})M_\eta^2\right)\\
&\qquad\times\left(3\beta_{\pi K}t+2(1-\beta_{\pi K})(M_K^2+M_\pi^2)+4(\alpha_{\pi K}-1)M_\pi M_K\right),
\end{split}\\
Z^{\eta\eta}_{\pi\eta}&=-4\alpha_{\pi\eta}\alpha_{\eta\eta}M_\pi^2\left(M_\pi^2-4M_\eta^2\right),\\
Z^{\pi\pi}_{\pi\eta}(t)&=4\alpha_{\pi\eta}M_\pi^2\left(6\beta_{\pi\pi}t+(5\alpha_{\pi\pi}-8\beta_{\pi\pi})M_\pi^2\right),\\
Y_{\pi\eta}^{\pi\eta}&=\alpha_{\pi\eta}^2 M_\pi^4,\\
Y^{KK}_{\pi\eta}(s)&=\frac38\left(3\beta_{\eta\pi
K}s-2(1+\beta_{\eta\pi K})M_K^2+(1-\alpha_{\eta\pi K}-\beta_{\eta\pi
K})M_\pi ^2+(1-\beta_{\eta\pi K})M_\eta^2\right)^2.
\end{align}

\

$\bullet\quad\pi^+\pi^-\rightarrow\pi^0\pi^0$
\begin{equation}\begin{split}
A_{\pi\pi}&=\frac1{3F_\pi^2}\left(\beta_{\pi\pi}\left(3s-4M_\pi^2\right)+\alpha_{\pi\pi}M_\pi^2\right)+\frac1{F_\pi^4}\left(\delta_{\pi\pi} s^2+\varepsilon_{\pi\pi}(t^2+u^2)\right)\\
&\quad+\frac1{72F_\pi^4} \left(Z_{\pi\pi}^{\pi\pi}(s)\overline{J}_{\pi\pi}(s)+Z^{\eta\eta}_{\pi\pi}\overline{J}_{\eta\eta}(s)
+Z^{KK}_{\pi\pi}(s)\overline{J}_{KK}(s)\right)\\
&\quad+\left[\frac1{72F_\pi^4}\left(Y^{KK}_{\pi\pi}(t,u)\overline{J}_{KK}(t)+Y_{\pi\pi}^{\pi\pi}(t,u)\overline{J}_{\pi\pi}(t)\right)\right]+[t\leftrightarrow u]+\Op5.
\end{split}\end{equation}
\begin{align}
Z_{\pi\pi}^{\pi\pi}(s)&=4\left(3\beta_{\pi\pi}s+(7\alpha_{\pi\pi}-4\beta_{\pi\pi})M_\pi^2\right) \left(3\beta_{\pi\pi}s+(\alpha_{\pi\pi}-4\beta_{\pi\pi})M_\pi^2\right),\\
Z^{\eta\eta}_{\pi\pi}&=4\alpha_{\pi\eta}^2M_\pi^4,\\
Z^{KK}_{\pi\pi}(s)&=\left(3\beta_{\pi K}s+2(1-\beta_{\pi K})(M_K^2+M_\pi^2)+4(\alpha_{\pi K}-1)M_\pi M_K\right)^2,\\
Y^{KK}_{\pi\pi}(t,u)&=3\gamma_{\pi K}^2(t-4M_K^2)(4M_\pi^2-t-2u),\\
Y_{\pi\pi}^{\pi\pi}(t,u)&=4\left(3\beta_{\pi\pi}^2t(t-u)+6\beta_{\pi\pi}(2\beta_{\pi\pi}u-\alpha_{\pi\pi}t)M_\pi^2+2(\alpha_{\pi\pi}^2+4\beta_{\pi\pi}(\alpha_{\pi\pi}-2\beta_{\pi\pi}))M_\pi^4\right).
\end{align}
\

$\bullet\quad\Ka K^0\rightarrow\eta\eta$
\begin{equation}\begin{split}
A_{\eta K}&=\frac1{4F_\pi^2}\left(\beta_{\eta K}(3s-2M_K^2-2M_\eta^2)+\alpha_{\eta K}\left(2M_\eta^2-\frac 23M_K^2\right)\right) +\frac1{4F_\pi^4}\left(\delta_{\eta K} s^2+\varepsilon_{\eta K}(t^2+u^2)\right)\\
&\quad+\frac1{24F_\pi^4}\left(Z_{\eta K}^{\eta\eta}(s)\overline{J}_{\eta\eta}(s)
+Z_{\eta K}^{\pi\pi}(s)\overline{J}_{\pi\pi}(s)+Z_{\eta K}^{KK}(s)\overline{J}_{KK}(s)
\right)\\
&\quad+\left[\frac{1}{32F_\pi^4}\left(Y_{\eta K}^{\eta K}(t,u)\overline{J}_{\eta K}(t)+Y_{\eta K}^{K\pi}(t,u)\overline{J}_{K \pi}(t)\right)\right. \\
&\qquad\quad-\frac{3}{32F_\pi^4}\frac1t\left(X_{\eta K}^{\eta K}(u)\overline{J}_{\eta K}(t)+X_{\eta K}^{K\pi}(u)\overline{J}_{K \pi}(t)\right)\\
&\quad\qquad\qquad+\left.\frac{3}{16F_\pi^4}\frac1{t^2}\Delt{K\eta}^2\left(W_{\eta K}^{\eta K}\overline{\overline{J}}_{\eta K}(t)+W_{\eta K}^{K\pi}\overline{\overline{J}}_{K \pi}(t)\right)\right]+\left[t\leftrightarrow u\right]+\Op5.
\end{split}\end{equation}
{\allowdisplaybreaks
\begin{align}
W_{\eta K}^{\eta K}&=\Delt{K\eta}^2\beta^2_{\eta K},\\
W_{\eta K}^{K\pi}&=\Delt{K\pi}^2\beta^2_{\eta\pi K},\\
X_{\eta K}^{\eta K}(u)&=\left(\beta_{\eta K}u+2\left(\beta_{\eta K}-\frac23\alpha_{\eta K}\right)M_K^2+2(\beta_{\eta K}+2\alpha_{\eta K})M_\eta^2 \right)\beta_{\eta K}\Delt{K\eta}^2,\\
\begin{split}
X_{\eta K}^{K\pi}(u)&=\beta_{\eta\pi K}\Big(\beta_{\eta\pi K}\left(M_K^2-M_\pi^2\right)^2 u+2\left(\beta_{\eta\pi K}-2\right)M_K^6 -2\left(2\beta_{\eta\pi K}-3\right)M_K^4 M_\eta^2\\
&\qquad+2\left(3+\beta_{\eta\pi K}-\alpha_{\eta\pi K}\right)M_K^4 M_\pi^2-2\left(\beta_{\eta\pi K}+1-\alpha_{\eta\pi K}\right)M_\pi^4 M_K^2 -2M_\eta^4 M_K^2\\
&\qquad\quad+2\left(\alpha_{\eta\pi K}-4\right)M_K^2 M_\eta^2 M_\pi^2-2\left(\alpha_{\eta\pi K}-1\right) M_\pi^4 M_\eta^2+2(\beta_{\eta\pi K}+1)M_\eta^4 M_\pi^2\Big),
\end{split}\\
Z_{\eta K}^{\eta\eta}(s)&=\left(3\beta_{\eta K}s-2(\beta_{\eta
K}+\frac13\alpha_{\eta K})M_K^2-2(\beta_{\eta K}-\alpha_{\eta K})
M_\eta^2\right)\alpha_{\eta\eta}\left(4M_\eta^2-M_\pi^2\right),\\
Z_{\eta K}^{\pi\pi}(s)&=\alpha_{\pi\eta}M_\pi^2\left(3\beta_{\pi K}s+2(1-\beta_{\pi K})(M_K^2+M_\pi^2)+4(\alpha_{\pi K}-1)M_\pi M_K\right),\\
\begin{split}
Z_{\eta K}^{KK}(s)&=3\left(3\beta _{\eta K}s-2\left(\beta_{\eta K}+\frac13\alpha_{\eta K}\right)M_K^2-2(\beta _{\eta K}-\alpha _{\eta K})M_\eta ^2\right)\\
&\qquad\times\left(\frac{3}{2}\beta_{KK}s+2(\alpha_{KK}-\beta_{KK})M_K^2\right),
\end{split}\\
\begin{split}
Y_{\eta K}^{\eta K}(t,u)&=\frac{1}{9}\Big(27\beta_{\eta K}^2t(t-u)-6\left(8\alpha_{\eta K}^2 - 8\alpha_{\eta K}\beta_{\eta K}+ 39\beta_{\eta K}^2 \right)M_K^2 M_{\eta}^2\\
&\qquad+ 9\left(8\alpha_{\eta K}^2 + 8\alpha_{\eta K}\beta_{\eta K}+5\beta_{\eta K}^2\right)M_{\eta}^4+
\left(8\alpha_{\eta K}^2 - 24\alpha_{\eta K}\beta_{\eta K} + 45\beta_{\eta K}^2\right)M_K^4
\\
&\quad\qquad+36\alpha_{\eta K}\beta_{\eta K}\left(M_K^2-3M_{\eta}^2\right)t+
54\beta_{\eta K}^2\left( M_K^2+M_{\eta}^2\right)u\Big),
\end{split}\\
\begin{split}
Y_{\eta K}^{K\pi}(t,u)&=3\beta_{\eta\pi K}^2 t(t-u)+\left(5\beta_{\eta\pi K}^2-8\beta_{\eta\pi K}+8\right)M_K^4 +2\left(\beta_{\eta\pi K}^2 +\beta_{\eta\pi K}+1\right)M_\eta^4\\
&\quad-8\left(2\beta_{\eta\pi K}^2+1\right)M_K^2 M_\eta^2+\left(2\beta_{\eta\pi K}\left(1-\alpha_{\eta\pi K}\right)+2\left(1-\alpha_{\eta\pi K}\right)^2-\beta_{\eta\pi K}^2\right)M_\pi^4\\
&\quad+2\left(2\left(\beta_{\eta\pi K}^2+\beta_{\eta\pi K}+1-\alpha_{\eta\pi K}\right)-\alpha_{\eta\pi K}\beta_{\eta\pi K}\right)M_\eta^2 M_\pi^2  \\
&\quad-2\left(5\beta_{\eta\pi K}^2+2\alpha_{\eta\pi K}\beta_{\eta\pi K}+4\left(1-\alpha_{\eta\pi K}\right)\right)M_K^2 M_\pi^2\\
&\quad-6\beta_{\eta\pi K}\left(M_\eta^2-2M_K^2-\left(\alpha_{\eta\pi K}-1\right)M_\pi^2\right)t+6\beta_{\eta\pi K}^2\left(M_K^2+M_\pi^2\right)u.
\end{split}
\end{align}
}
\

$\bullet\quad\Ka K^0\rightarrow\pi^0\eta$
\begin{equation}\begin{split}
A_{\eta\pi K}&=\frac{1}{4\sqrt{3}F_\pi ^2}V_{\eta\pi K}(s)+\frac{1}{4\sqrt{3}F_\pi^4}\left(\delta_{\eta\pi K}s^2+\varepsilon_{\eta\pi K}(t^2+u^2) \right)\\
&\quad+\frac1{24\sqrt{3}F_\pi^4}\left(Z_{\eta\pi K}^{\eta\pi}(s)\overline{J}_{\eta\pi}(s)+
Z_{\eta\pi K}^{KK}\overline{J}_{KK}(s)\right) V_{\eta\pi K}(s)\\
&\quad+\Big[-\frac{1}{96\sqrt{3}F_\pi^4}\left(Y_{\eta\pi K}^{\eta K}(t,u)\overline{J}_{\eta K}(t)+Y_{\eta\pi K}^{K\pi}(t,u)\overline{J}_{K \pi}(t)\right)\\
&\quad\qquad-\frac{1}{32\sqrt{3}F_\pi^4}\frac1t\left(X_{\eta\pi K}^{\eta K}(u)\overline{J}_{\eta K}(t)+X_{\eta\pi K}^{K\pi}(u)\overline{J}_{K \pi}(t)\right)\\
&\quad\qquad\qquad+\frac{1}{16\sqrt{3}F_\pi^4}\frac1{t^2}U_{\eta\pi K}\left.\left(W_{\eta\pi K}^{\eta K}\overline{\overline{J}}_{\eta K}(t) +W_{\eta\pi K}^{K\pi}\overline{\overline{J}}_{K \pi}(t)\right)\right]+[t\leftrightarrow u]+\Op5.
\end{split}\end{equation}

{\allowdisplaybreaks
\begin{align}
U_{\eta\pi K}&=\beta_{\eta\pi K} \Delt{K\eta}\Delt{K\pi},\\
V_{\eta\pi K}(s)&=\beta_{\eta\pi K}\left(3s-2M_K^2-M_\pi ^2-M_\eta^2\right)-\left( 2M_K^2-M_\pi ^2-M_\eta ^2+\alpha_{\eta\pi K}M_\pi ^2\right),\\
Z_{\eta\pi K}^{\eta\pi}&=2\alpha_{\pi\eta}M_\pi^2,\\
Z_{\eta\pi K}^{KK}(s)&=\frac32\beta_{KK}s+2(\alpha_{KK}-\beta_{KK})M_K^2,\\
\begin{split}
Y_{\eta\pi K}^{\eta K}(t,u)&=9\beta_{\eta\pi K}\beta_{\eta K}(t(u-t)-2(M_\eta^2+M_K^2)u)-3(\beta_{\eta\pi K}+2)(2\alpha_{\eta K}+\beta_{\eta K})M_\eta^4\\
&\quad+\left(3\beta_{\eta K}(4-5\beta_{\eta\pi K} ) +4\alpha_{\eta K}(\beta_{\eta\pi K}-2)\right)M_K^4\\
&\quad+2(\alpha_{\eta K}(2-2\alpha_{\eta\pi K}+\beta_{\eta\pi K})+3\beta_{\eta K}(\alpha_{\eta\pi K}-1+4\beta_{\eta\pi K}))M_K^2 M_\pi^2\\
&\quad+(\alpha_{\eta K}(28-10\beta_{\eta\pi K})+6\beta_{\eta K}(9\beta_{\eta\pi K}+1))M_K^2 M_\eta^2 \\
&\quad+6((\alpha_{\eta\pi K}-1)(2\alpha_{\eta K}+\beta_{\eta K})-\beta_{\eta\pi K}(\alpha_{\eta K}+2\beta_{\eta K}))M_\eta^2 M_\pi^2\\
&\quad-3t\big(2(3\beta_{\eta K}+\alpha_{\eta K}\beta_{\eta\pi K})
M_K^2-3(\beta_{\eta K}+2\alpha_{\eta K}\beta_{\eta\pi
K})M_\eta^2+3\beta_{\eta K}(\alpha_{\eta\pi K}-1)M_\pi^2\big),
\end{split}\\
\begin{split}
Y_{\eta\pi K}^{K\pi}(t,u)&=3\beta_{\eta\pi K}t\left((10\gamma_{\pi K}-\beta_{\pi K})t+(\beta_{\pi K} +2\gamma_{\pi K})u\right) +3(\beta_{\pi K}-2\gamma_{\pi K}(3+2\beta_{\eta\pi K}))tM_\eta^2\\
&\quad-6\beta_{\eta\pi K} (\beta_{\pi K}+2\gamma_{\pi K})(M_K^2+M_\pi^2)u-6(2\gamma_{\pi K}(4\beta_{\eta\pi K}-3) +\beta_{\pi K}-\beta_{\eta\pi K})tM_K^2\\
&\quad-3(2\beta_{\eta\pi K}(6\gamma_{\pi K}-1)+(1-\alpha_{\eta\pi K})(6\gamma_{\pi K}-\beta_{\pi K})) tM_\pi^2\\
&\quad+(2+\beta_{\eta\pi K}-2\alpha_{\eta\pi K})(6\gamma_{\pi K}-\beta_{\pi K}-2)M_\pi^4\\
&\quad+\left(\beta_{\pi K}(4-5\beta_{\eta\pi K}) +2\gamma_{\pi K}(21\beta_{\eta\pi K}-12)+8-4\beta_{\eta\pi K} \right)M_K^4\\
&\quad+2(\beta_{\eta\pi K}(4\beta_{\pi K}-6\gamma_{\pi K}-1)-\beta_{\pi K}+6\gamma_{\pi K}-2)M_K^2 M_\eta^2\\
&\quad-2(\beta_{\pi K}(2\beta_{\eta\pi K}+1)-6\gamma_{\pi K}(3\beta_{\eta\pi K}+1)+2+\beta_{\eta\pi K})M_\pi^2 M_\eta^2\\
&\quad+4(\alpha_{\pi K}-1)(V_{\eta\pi K}(t)+3(2M_K^2-M_\pi^2-M_\eta^2+\alpha_{\eta\pi K}M_\pi^2))M_K M_\pi\\
&\quad+2(3\beta_{\eta\pi K}(3\beta_{\pi K}+4\gamma_{\pi K}-1)+(\alpha_{\eta\pi K}+1)(\beta_{\pi K} -6\gamma_{\pi K}+2))M_K^2 M_\pi^2,
\end{split}\\
\begin{split}
X_{\eta\pi K}^{\eta K}(u)&=\Delt{\eta K}\Big(3\beta_{\eta K}\beta_{\eta\pi K}(M_\eta^2-M_K^2)u+3\beta_{\eta K}M_\eta^4+(2\alpha_{\eta K}\beta_{\eta\pi K}+6\beta_{\eta K}\left(1-\beta_{\eta\pi K}\right)) M_K^4\\
&\qquad\quad+3(\beta_{\eta K}(1-\alpha_{\eta\pi K})+2\beta_{\eta \pi K}(\alpha_{\eta K}+\beta_{\eta K})) M_\eta^2 M_\pi^2\\
&\qquad\quad\quad+(3\beta_{\eta K}(\alpha_{\eta\pi K}-1)+2\beta_{\eta \pi K}(3\beta_{\eta K}-\alpha_{\eta K})) M_K^2 M_\pi^2\\
&\qquad\quad\quad\quad-3(3\beta_{\eta K}+2\beta_{\eta\pi K}(\beta_{\eta K}+\alpha_{\eta K}))M_K^2 M_\eta^2 \Big),
\end{split}\\
\begin{split}
X_{\eta\pi K}^{K\pi}(u)&=\Delt{\pi K}\Big(\beta_{\eta\pi K}(\beta_{\pi K}+2\gamma_{\pi K})\left(M_\pi^2-M_K^2\right)u +4\beta_{\eta\pi K}(1-\alpha_{\pi K})(M_K^2-M_\eta^2)M_K M_\pi\\
&\qquad\quad+2\beta_{\eta\pi K}(2\gamma_{\pi K}-\beta_{\pi K}-1)(M_K^4-M_K^2 M_\eta^2-M_\pi^2 M_\eta^2+M_K^2 M_\pi^2)\\
&\qquad\quad\quad+(\beta_{\pi K}+2\gamma_{\pi K})(2M_K^4-M_K^2 M_\eta^2+M_\pi^2 M_\eta^2+(\alpha_{\eta\pi K}-3)M_K^2 M_\pi^2)\\
&\qquad\quad\quad\quad+(\beta_{\pi K}+2\gamma_{\pi K})(1-\alpha_{\eta\pi K})M_\pi^4\Big),
\end{split}\\
W_{\eta\pi K}^{\eta K}&=3\Delt{K\eta}^2\beta_{\eta K},\\
W_{\eta\pi K}^{K\pi}&=\Delt{K\pi}^2(\beta_{\pi K}+2\gamma_{\pi K}).
\end{align}}%

\

$\bullet\quad\pi^0\pi^0\rightarrow\Ka K^0$
\begin{equation}\begin{split}
A_{\pi K}&=\frac1{12F_\pi^2}\left(\beta_{\pi K}\left(3s-2M_K^2-2M_\pi^2\right)+2(M_K-M_\pi)^2+4\alpha_{\pi K}M_\pi M_K\right)\\
&\quad+\frac1{12F_\pi^4}\left(\delta_{\pi K}s^2+\varepsilon_{\pi K} (t^2+u^2)\right)\\
&\quad+\frac1{72F_\pi^4}\Big(Z_{\pi K}^{\pi\pi}(s)\overline{J}_{\pi\pi}(s)+Z_{\pi K}^{\eta\eta}(s)\overline{J}_{\eta\eta}(s)+Z_{\pi K}^{KK}(s) \overline{J}_{KK}(s)\Big)\\
&\quad+\bigg[\frac{1}{96F_\pi^4}\left(Y_{\pi K}^{\eta K}(t,u)\overline{J}_{\eta K}(t)+Y_{\pi K}^{\pi K}(t,u)\overline{J}_{\pi K}(t)\right)\\
&\quad\qquad+\frac{1}{32F_\pi^4}\frac1t\left(X_{\pi K}^{\eta K}(u)\overline{J}_{\eta K}(t)+X_{\pi K}^{\pi K}(u)\overline{J}_{\pi K}(t)\right)\\
&\quad\qquad\qquad+\frac{1}{16F_\pi^4}\frac1{t^2}\Delt{K\pi}^2\left(W_{\pi K}^{\eta K}\overline{\overline{J}}_{\eta K}(t)
+W_{\pi K}^{\pi K}\overline{\overline{J}}_{\pi K}(t)\right)\bigg]+[t\leftrightarrow u]+\Op5.
\end{split}\end{equation}

{\allowdisplaybreaks
\begin{align}
\begin{split}
Z_{\pi K}^{\pi\pi}(s)&=\left(3\beta_{\pi K}s+2(1-\beta_{\pi K})(M_K^2+M_\pi^2)+4(\alpha_{\pi K}-1)M_\pi M_K\right)\\
&\qquad\times\left(6\beta_{\pi\pi}s+(5\alpha_{\pi\pi} -8\beta_{\pi\pi})M_\pi^2\right),
\end{split}\\
Z_{\pi K}^{\eta\eta}(s)&=3\alpha_{\pi\eta}M_\pi^2\left(3\beta_{\eta K}s-2\left(\beta_{\eta K}+\frac13\alpha_{\eta K}\right)M_K^2+2(\alpha_{\eta K}-\beta _{\eta K})M_\eta^2\right),\\
\begin{split}
Z_{\pi K}^{KK}(s)&=3\left(3\beta_{\pi K}s+2(1-\beta_{\pi K})(M_K^2+M_\pi^2)+4(\alpha_{\pi K}-1)M_\pi M_K\right)\\
&\qquad\times\left(\frac{3}{2}\beta_{KK}s+2(\alpha _{KK}-\beta_{KK})M_K^2\right),
\end{split}\\
\begin{split}
Y_{\pi K}^{\eta K}(t,u)&=3\beta_{\eta K}^2t(t-u)-6\beta_{\eta\pi K}\left(M_\eta^2-2M_K^2+(1-\alpha_{\eta\pi K})M_\pi^2\right)t +6\beta_{\eta\pi K}^2(M_\eta^2+ M_K^2)u\\
&\quad+\left(5\beta_{\eta\pi K}^2 + 8(1-\beta_{\eta\pi K})\right)M_K^4 +\left(2+2\beta_{\eta\pi K}-\beta_{\eta\pi K}^2\right)M_\eta^4 -2(5\beta_{\eta\pi K}^2+4)M_K^2 M_\eta^2 \\
&\quad+2\left(\alpha_{\eta\pi K}(\alpha_{\eta\pi K}-\beta_{\eta\pi K}-2)+\beta_{\eta\pi K}^2+\beta_{\eta\pi K}+1\right)M_\pi^4\\
&\quad-4(\beta_{\eta\pi K}(4\beta_{\eta\pi K}+\alpha_{\eta\pi K})+2(1-\alpha_{\eta\pi K}))M_K^2 M_\pi^2\\
&\quad+2(2(\beta_{\eta\pi K}^2+\beta_{\eta\pi K}+1)-\alpha_{\eta\pi K}(\beta_{\eta\pi K}+2))M_\pi^2 M_\eta^2,
\end{split}\\
\begin{split}
Y_{\pi K}^{\pi K}(t,u)&=\frac{1}{3}\Big(3t\left(\left(\beta_{\pi K}^2+26\gamma_{\pi K}^2\right)t-\left(\beta_{\pi K}^2+2\gamma_{\pi K}^2\right)u\right)+6\left(\beta_{\pi K}^2+ 2 \gamma_{\pi K}^2\right)\left(M_K^2+M_\pi^2\right)u\\
&\qquad-12\left(\left(8\gamma_{\pi K}^2+\beta_{\pi K}\right)\left(M_K^2+M_\pi^2\right) +2\beta_{\pi K}\left(\alpha_{\pi K}-1\right)M_K M_\pi\right)t \\
&\qquad\quad+2\left(16\alpha_{\pi K}\left(\alpha_{\pi K}-2\right)+\beta_{\pi K} \left(8-13\beta_{\pi K}\right)+6\left(4+13\gamma_{\pi K}^2\right)\right)M_K^2 M_\pi^2\\
&\qquad\quad\quad+\left(5\beta_{\pi K}^2+8\left(\beta_{\pi K}+1\right)-30\gamma_{\pi K}^2\right)\left(M_K^4+M_\pi^4\right)\\
&\qquad\quad\quad\quad+16\left(\alpha_{\pi K}-1\right)\left(\beta_{\pi K}+2\right)M_K M_\pi\left(M_K^2+M_\pi^2\right)\Big),
\end{split}\\
\begin{split}
X_{\pi K}^{\eta K}(u)&=\beta_{\eta\pi K}\Big(-\beta_{\eta\pi K}\Delt{K\eta}^2u+2\left(2-\beta_{\eta\pi K}\right)M_K^6+2M_\eta^4\left(\left(\beta_{\eta\pi K}+1\right)M_K^2 -M_\pi^2\right)\\
&\qquad+2M_K^2 M_\pi^2 M_\eta^2 \left(4-\alpha_{\eta\pi K}\right)+2M_K^4\left(\left(\alpha_{\eta\pi K}+2\beta_{\eta\pi K}-3\right)M_\pi^2-\left(\beta_{\eta\pi K}+3\right)M_\eta^2\right)\\
&\quad\qquad+2M_\pi^4\left(\left(\alpha_{\eta\pi K}-\beta_{\eta\pi K}-1\right)M_\eta^2+\left(1-\alpha_{\eta\pi K}\right)M_K^2\right)\Big),
\end{split}\\
\begin{split}
X_{\pi K}^{\pi K}(u)&=-\frac13\Delt{K\pi}^2 \Big(\left(\beta_{\pi K}^2+2\gamma_{\pi K}^2\right)u+8\left(\alpha_{\pi K}-1\right) \beta_{\pi K}M_K M_\pi \\
&\qquad\qquad\qquad\qquad\qquad\qquad\qquad\qquad+2 \left(\beta_{\pi K}^2+2\beta_{\pi K}-6\gamma_{\pi K}^2\right) \left(M_K^2+M_\pi^2\right)\Big),
\end{split}\\
W_{\pi K}^{\eta K}&=\Delt{K\eta}^2\beta^2_{\eta\pi K},\\
W_{\pi K}^{\pi K}&=\frac13\Delt{K\pi}^2\left(\beta^2_{\pi K}+2\gamma_{\pi K}^2\right).
\end{align}}%

\

$\bullet\quad\pi^-\pi^0\rightarrow K^- K^0$
\begin{equation}\begin{split}
B_{\pi K}&=\frac{1}{4F_\pi^2}\gamma_{\pi K}(t-u)+\frac{1}{4F_\pi^4}\varphi_{\pi K} s(t-u)\\
&\quad+\frac{1}{48F_\pi^4}\gamma_{\pi K}(t-u)\left(2\overline{J}_{\pi\pi}(s)\beta_{\pi\pi}(s-4M_\pi^2)+\overline{J}_{KK}(s)\beta_{KK}(s-4M_K^2)\right)\\
&\quad+\bigg[\frac{1}{96F_\pi^4}\left(Y_{\pi Kch}^{\eta K}(t,u)\overline{J}_{\eta K}(t)+Y_{\pi Kch}^{\pi K}(t,u)\overline{J}_{\pi K}(t)\right)\\
&\quad\qquad+\frac{1}{32F_\pi^4}\frac1t\left(X_{\pi Kch}^{\eta K}(u)\overline{J}_{\eta K}(t)+X_{\pi Kch}^{\pi K}(u)\overline{J}_{\pi K}(t)\right)\\
&\quad\qquad\qquad+\frac{1}{16F_\pi^4}\frac1{t^2}\Delt{K\pi}^2\left(W_{\pi Kch}^{\eta K}\overline{\overline{J}}_{\eta K}(t) +W_{\pi Kch}^{\pi K}\overline{\overline{J}}_{\pi K}(t)\right)\bigg]-[t\leftrightarrow u]+\Op5.
\end{split}\end{equation}

\begin{align}
Y_{\pi Kch}^{\eta K}(t,u)&=Y_{\pi K}^{\eta K}(t,u), \\
\begin{split}
Y_{\pi Kch}^{\pi K}(t,u)&=-\gamma_{\pi K}\Big(t\left((10\beta_{\pi K}-13\gamma_{\pi K})t+(2\beta_{\pi K}+\gamma_{\pi K})u\right)\\
&\qquad\quad-4((4(\beta_{\pi K}-\gamma_{\pi K})+3)(M_K^2+M_\pi^2)+6(\alpha_{\pi K}-1)M_K M_\pi)t\\
&\qquad\quad\quad-2(2\beta_{\pi K}+\gamma_{\pi K})(M_K^2+M_\pi^2)u+(14\beta_{\pi K}+5\gamma_{\pi K}+8)(M_K^4+M_\pi^4)\\
&\qquad\quad\quad\quad+2(2\beta_{\pi K}-13\gamma_{\pi K}+8)M_K^2 M_\pi^2+16(\alpha_{\pi K}-1)M_K M_\pi (M_K^2+M_\pi^2)\Big),
\end{split}\\
X_{\pi Kch}^{\eta K}(u)&=X_{\pi K}^{\eta K}(u),\\
\begin{split}
X_{\pi Kch}^{\pi K}(u)&=-\frac13\gamma_{\pi K}(M_K^2-M_\pi^2)^2\Big(\left(2\beta_{\pi K}+\gamma_{\pi K} \right)u+8\left(\alpha_{\pi K}-1\right)M_K M_\pi\\
&\qquad\qquad\qquad\qquad\qquad\qquad\qquad+2\left(2-2\beta_{\pi K}-3\gamma_{\pi K}\right) \left(M_K^2+M_\pi^2\right)\Big),
\end{split}\\
W_{\pi Kch}^{\eta K}&=W_{\pi K}^{\eta K},\\
W_{\pi Kch}^{\pi K}&=\frac13\gamma_{\pi K}\Delt{K\pi}^2\left(2\beta_{\pi K}+\gamma_{\pi K}\right).
\end{align}

\

$\bullet\quad K^- K^+\rightarrow \Ka K^0$
\begin{equation}\begin{split}
A_{KK}&=\frac1{6F_\pi ^2}\left(\beta_{KK}\left(4M_K^2-3u\right)+3\gamma_{KK}(s-t)+2\alpha _{KK}M_K^2\right)+\frac1{6F_\pi^4}\left(\delta_{KK} s^2 +\varepsilon_{KK} t^2+\varphi_{KK} st\right)\\
&\quad+\frac1{288F_\pi^4}\Big(Z_{KK}^{KK}(s,t)\overline{J}_{KK}(s)+Z^{\pi\eta}_{KK}(s)\overline{J}_{\pi\eta}(s)+Z^{\eta\eta}_{KK}(s)\overline{J}_{\eta\eta}(s) +Z_{KK}^{\pi\pi}(s,t)\overline{J}_{\pi\pi}(s)\Big)\\
&\quad+\frac1{72F_\pi^4}\left(Y_{KK}^{KK}(s,t)\overline{J}_{KK}(t)+Y_{KK}^{\pi\eta}(t)\overline{J}_{\pi\eta}(t)+Y_{KK}^{\pi\pi}(s,t)\overline{J}_{\pi\pi}(t)\right)\\
&\quad+\frac1{36F_\pi^4}\overline{J}_{KK}(u)\left(3\beta_{KK}u-2(\alpha_{KK}+2\beta_{KK})M_K^2\right)^2+\Op5.
\end{split}\end{equation}

\begin{align}
Z_{KK}^{KK}(s,t)&=8\Big(\left(3\beta_{KK}s+4(\alpha_{KK}-\beta_{KK})M_K^2\right)^2+3\beta_{KK}^2(s-4M_K^2)(s+2t-4M_K^2)\Big)\\
Z^{\pi\eta}_{KK}(s)&=-6\Big(3\beta_{\eta\pi K}s-2(1+\beta_{\eta\pi K})M_K^2+(1-\alpha_{\eta\pi K}-\beta_{\eta\pi K})M_\pi^2+(1-\beta_{\eta\pi K})M_\eta^2\Big)^2\\
Z^{\eta\eta}_{KK}(s)&=\left(9\beta_{\eta K}s-2(3\beta_{\eta K}+\alpha_{\eta K})M_K^2+6(\alpha_{\eta K}-\beta_{\eta K})M_\eta^2\right)^2\\
\begin{split}
Z_{KK}^{\pi\pi}(s,t)&=3\Big(s\left((9\beta_{\pi K}^2-2\gamma_{\pi K}^2)s-4\gamma_{\pi K}^2t\right)+4(1-\beta_{\pi K})^2(M_K^4+M_\pi^4)\\
&\qquad+4s\left(3\beta_{\pi K}(1-\beta_{\pi K})+2\gamma_{\pi K}^2\right)(M_K^2+M_\pi^2)\\
&\qquad\quad+8(2\gamma_{\pi K}^2 t+((1-\beta_{\pi K})^2+2(1-\alpha_{\pi K})^2-4\gamma_{\pi K}^2)M_K^2)M_\pi^2\\
&\qquad\quad\quad+8(\alpha_{\pi K}-1)(3\beta_{\pi K} s +2(1-\beta_{\pi K})(M_K^2+M_\pi^2))M_\pi M_K\Big)
\end{split}\\
Y_{KK}^{KK}(s,t)&=\frac1{16}Z_{KK}^{KK}(t,s)\\
Y_{KK}^{\pi\eta}(t)&=3\Big(3\beta_{\eta\pi K}t-2(1+\beta_{\eta\pi K})M_K^2+(1-\alpha_{\eta\pi K}-\beta_{\eta\pi K})M_\pi^2+(1-\beta_{\eta\pi K})M_\eta^2\Big)^2\\
Y_{KK}^{\pi\pi}(s,t)&=3\gamma_{\pi K}^2(2s+t-4M_K^2)(t-4M_\pi^2)
\end{align}

\section{Standard chiral perturbation theory \Op4 values of the polynomial parameters}
In this appendix we give the values of our polynomial parameters which reproduce the \Op4 results of the standard chiral perturbation theory. To that end we have used the computation of \cite{Spanele}, which contains all the considered amplitudes and its advantage is also that their results are in the unitary form.

The form of our results is in terms of physical observables and thus has to be scale-independent and the same for all the possible regularization schemes. The only thing which can change by an eventual change of the scheme is the relation between the parameters of our parametrization and
the (renormalised) constants of the Lagrangian theory (LEC). By the change of the scale, the values of the LEC change, but their combinations giving the value of our parameters remain scale-independent - what can be another test of the results obtained from the Lagrangian theory. A further difference between (Lagrangian theory) results of different authors can rise from a different choice of the way they parameterize the \Op2 constants (bare masses and decay constants) using the physical parameters, e.g.\ in \cite{Spanele} they expand $F_K$ and $F_\eta$ decay constants in terms of $F_\pi$, $L_4^r$ and $L_5^r$.

In \cite{Spanele} they have used the Gell-Mann-Okubo relation (GMO) to get their results more simplified. In the standard chiral power counting the GMO formula has correction of the \Op4 order and thus we could also use it to simplify our results in some places, where it would give only correction of the \Op6 order at least. However, to let these relations be closely connected to the results \cite{Spanele}, we have not done it and the only place where we refer to the GMO formula (and its \Op4 order correction) is in those places where we want to emphasize the validity of the \Op2 values of the parameters from Appendix~A. Nevertheless, the use of
\begin{equation}
\Delta_{\mathrm{GMO}}(M_\eta^2-M_\pi^2)=4M_K^2-M_\pi^2-3M_\eta^2,
\end{equation}
which is in the standard power counting of the \Op4 order, can also be understood just as a (more complicated) notation of the right-hand side of this definition.

Other objects appearing in the relations are the chiral logarithms, given in accordance with \cite{Spanele} by
\begin{equation}
\mu_i=\frac{M_i^2}{32\pi^2 F_\pi^2}\log\frac{M_i^2}{\mu^2} \qquad \text{with }i=\pi, K, \eta.
\end{equation}
(As has been already stated, its dependence on the scale $\mu$ is compensated on the right-hand side of the following relations by the scale-dependence of LEC $L_i^r$ as listed in \cite{Gasser}.)

By the comparison of \cite{Spanele} with our results from the previous appendix we get the following relations:

{\allowdisplaybreaks
$\bullet\quad \eta\eta$
\begin{align}
\delta _{\eta\eta}&=12(2L_1^r+2L_2^r+L_3)-\frac{27F_\pi ^2\mu _K}{4M_K^2}-%
\frac{27}{128\pi ^2}\,,\\
\begin{split}
F_\pi^2(\alpha_{\eta\eta}-1)(M_\pi^2-4M_\eta^2)&=-\frac43F_\pi^2\Delta_{\mathrm{GMO}}(M_\eta^2-M_\pi^2)+96M_\eta ^4(L_4^r-3L_6^r)
\\ &
+8L_5^r\left( 3M_\pi ^4-10M_\eta ^2M_\pi ^2+13M_\eta ^4\right)\\
&-%
192L_7\left( M_\pi ^4-3M_\eta ^2M_\pi ^2+2M_\eta ^4\right)-48L_8^r\left( 2M_\pi ^4-6M_\eta ^2M_\pi ^2+7M_\eta ^4\right)\\
&+\frac{\left( 7M_\pi^4-24M_\pi^2M_\eta ^2-54M_\eta^4\right) \mu _K F_\pi^2}{3M_K^2}+\frac{\left(
48M_\eta ^2-7M_\pi ^2\right) \mu _\pi F_\pi^2}{3}
\\
&+\frac{\left( M_\pi ^4-8M_\eta ^2M_\pi ^2+24M_\eta ^4\right) \mu _\eta F_\pi^2}{%
M_\eta ^2}+\frac{80M_K^4-176M_\eta
^2M_K^2+103M_\eta^4}{32\pi^2}\,.
\end{split}
\end{align}
$\bullet\quad \pi\eta$
\begin{align}
\varepsilon _{\pi \eta }&=24L_1^r+4L_3-\frac{9F_\pi ^2\mu _K}{4M_K^2}-\frac
9{128\pi ^2}\,,\\
\delta _{\pi \eta }&=12L_2^r+4L_3-\frac{9F_\pi ^2\mu _K}{4M_K^2}-\frac
9{128\pi ^2}\,,\\
F_\pi ^2\beta _{\pi \eta }&=8L_4^r \left( M_\pi
^2+M_\eta ^2\right)-\frac{M_\pi ^2+M_\eta ^2}{32\pi ^2}-%
\frac{\left( M_\pi ^2+3M_\eta ^2\right)F_\pi ^2 \mu _K}{3M_K^2}-\frac{2F_\pi ^2\mu _\pi }3\,,\\
\begin{split}
F_\pi ^2(\alpha _{\pi \eta }-1)M_\pi ^2 &=16L_4^r\left( M_\pi ^4-4M_\eta ^2M_\pi ^2+M_\eta ^4\right)+16M_\eta ^2M_\pi ^2(-L_5^r+6L_6^r)
\\ &
+96L_7\left( M_\pi ^2-M_\eta^2\right)M_\pi ^2+48L_8^rM_\pi^4\\
&-\frac{\left( 5M_\pi ^4-28M_\eta ^2M_\pi ^2+6M_\eta ^4\right) F_\pi ^2\mu _K}{%
3M_K^2}+\frac{\left( -17M_\pi ^4+9M_\eta ^2M_\pi ^2+4M_\eta
^4\right)F_\pi ^2 \mu _\pi }{3\left( M_\pi ^2-M_\eta ^2\right) } \\
&+\frac{\left( M_\pi ^4-M_\eta ^2M_\pi ^2+4M_\eta ^4\right)M_\pi^2F_\pi ^2 \mu_\eta }{%
3M_\eta^2(M_\pi ^2-M_\eta ^2)}-\frac{8M_\pi ^4-11M_\eta ^2M_\pi ^2+6M_\eta ^4}{96\pi ^2}\,.
\end{split}
\end{align}
$\bullet\quad \pi\pi$
\begin{align}
\varepsilon_{\pi \pi }&=4L_2^r+\frac 1{12}F_\pi ^2\left( -\frac{\mu _K}{M_K^2}-%
\frac{8\mu _\pi }{M_\pi ^2}\right) -\frac 7{384\pi ^2}\,,\\
\delta _{\pi \pi }&=8L_1^r+4L_3+\frac 1{12}F_\pi ^2\left( -\frac{\mu _K}{%
M_K^2}-\frac{8\mu _\pi }{M_\pi ^2}\right) -\frac{13}{384\pi ^2}\,,
\\
F_\pi ^2(\beta _{\pi \pi }-1)&=8\left(2L_4^r+L_5^r\right) M_\pi ^2-\frac{F_\pi ^2\mu _KM_\pi ^2}{%
M_K^2}-\frac{5M_\pi ^2}{32\pi ^2}-4F_\pi ^2\mu _\pi,\\
F_\pi ^2(\alpha _{\pi \pi }-1)M_\pi ^2&=-16\left(
2L_4^r+L_5^r-6L_6^r-3L_8^r\right) M_\pi ^4-%
\frac{F_\pi ^2\mu _\eta M_\pi ^4}{3M_\eta ^2}-\frac{F_\pi ^2\mu _K M_\pi ^4}{M_K^2}-\frac{7M_\pi ^4}{96\pi ^2}%
-F_\pi ^2\mu _\pi M_\pi^2.
\end{align}
$\bullet\quad \eta K$
\begin{align}
\varepsilon _{\eta K}&=16L_2^r+\frac{4}3L_3-\frac{F_\pi ^2(2\mu_K+\mu _\pi+3\mu_\eta) }{2\left( M_K^2-M_\eta ^2\right) }+\frac 1{64\pi ^2}\,,
\\
\delta _{\eta K}&=32L_1^r+\frac{40}3L_3+\frac{\left( 9M_\eta
^2-8M_K^2\right) F_\pi ^2\mu _K}{2M_K^2\left( M_K^2-M_\eta ^2\right) }+\frac{F_\pi^2\left(\mu _\pi -3\mu _\eta\right)}{4(M_K^2-M_\eta^2)}-\frac{11}{64\pi ^2}\,,\\
\begin{split}
F_\pi ^2(\beta _{\eta K}-1) &=\frac{32}{3}\left( M_K^2+M_\eta ^2\right)L_4^r+8L_5^rM_\pi^2-\frac{M_\eta^2\left(7M_K^2-9M_\eta ^2\right)F_\pi ^2 \mu _K}{3M_K^2(M_K^2-M_\eta^2)}\\
&+\frac{%
\left(13M_\eta ^2-11M_K^2\right) F_\pi ^2\mu _\pi }{3\left( M_K^2-M_\eta ^2\right) }
+\frac{\left(4M_K^4-15M_\eta ^2M_K^2+7M_\eta ^4\right) F_\pi ^2\mu _\eta }{%
3M_\eta ^2(M_K^2-M_\eta ^2)}
\\ &
-\frac{17M_\eta ^2+37M_K^2}{192\pi ^2}\,,
\end{split}\\
\begin{split}
F_\pi ^2(\alpha _{\eta K}-1)\left(3M_\eta^2-M_K^2\right) &=F_\pi^2\Delta_{\mathrm{GMO}}(M_\eta^2-M_\pi^2)+32L_4^r\left( M_K^4-4M_\eta
^2M_K^2+M_\eta ^4\right)\\
&+192L_6^rM_K^2M_\eta ^2+4L_5^r\left( -3M_\pi ^4+6M_K^2M_\pi ^2+M_\eta ^2M_\pi ^2-12M_\eta
^4\right)\\
&+48L_7\left( M_\pi ^4-4M_\eta ^2M_\pi ^2+3M_\eta ^4\right)+24L_8^r\left( M_\pi ^4-3M_\eta^2M_\pi ^2+6M_\eta ^4\right)
\\ &
-\frac{\left(14M_K^6-84M_\eta ^2M_K^4+102M_\eta ^4M_K^2-27M_\eta ^6\right) F_\pi ^2\mu _K}{3(
M_K^2-M_\eta ^2) }\\
&-\frac{\left( M_K^4+12M_\eta^2M_K^2-14M_\eta ^4\right)F_\pi ^2 \mu _\pi }{2(M_K^2-M_\eta ^2)}
\\ &
+\frac{\left( 8M_K^6-41M_\eta ^2M_K^4+104M_\eta ^4M_K^2-78M_\eta ^6\right)
F_\pi ^2\mu _\eta }{6M_\eta ^2(M_K^2-M_\eta^2)}\\
&-\frac{31M_K^4-55M_\eta ^2M_K^2+39M_\eta ^4}{96\pi ^2}\,.
\end{split}
\end{align}
$\bullet\quad \eta\pi K$
\begin{align}
\varepsilon _{\eta \pi K}&=-4L_3+\frac{6\mu _KF_\pi ^2}{M_K^2-M_\pi ^2}-\frac{3\mu
_\pi F_\pi ^2}{2\left( M_K^2-M_\pi ^2\right) }+\frac{3\mu _\eta F_\pi ^2}{%
2\left( M_K^2-M_\eta ^2\right) }+\frac 1{64\pi ^2}\,,
\\
\delta _{\eta \pi K}&=8L_3+\frac{3\left( M_\pi ^2-2M_K^2\right) \mu _KF_\pi ^2}{%
2M_K^2\left( M_K^2-M_\pi ^2\right) }-\frac{3\mu _\pi F_\pi ^2}{4\left(
M_K^2-M_\pi ^2\right) }-\frac{3\mu _\eta F_\pi ^2}{4\left( M_K^2-M_\eta
^2\right) }-\frac 5{64\pi ^2}\,,\\
\begin{split}
F_\pi ^2(\beta _{\eta \pi K}-1) &=\frac{4}{3}L_3\Delta_{\mathrm{GMO}}(M_\eta^2-M_\pi^2)+8M_\pi ^2L_5^r
\\ &
+\frac{\left(-32M_K^4+2M_K^2(2M_\pi^2-3M_\eta^2)+3M_\pi^2(M_\pi^2+M_\eta ^2)\right)F_\pi ^2 \mu _K}{6\left(
M_K^2-M_\pi ^2\right) }\\
&+\frac{\left(-40M_K^2+9M_\eta^2+33M_\pi^2\right)F_\pi ^2 \mu _\pi }{12\left( M_K^2-M_\pi ^2\right) }-\frac{\left(M_\pi ^2+5M_\eta ^2\right)F_\pi ^2 \mu _\eta }{4(M_K^2-M_\eta ^2)}+\frac{-57M_K^2+4M_\pi^2+27M_\eta ^2}{192\pi ^2}\,,
\end{split}\\
\begin{split}
F_\pi ^2\alpha _{\eta \pi K}M_\pi ^2 &=-\frac{F_\pi^2 }{3}\Delta_{\mathrm{GMO}}(M_\eta^2-M_\pi^2)+\frac{8}{3}L_3\left(8M_K^2+M_\pi ^2+M_\eta^2\right)\Delta_{\mathrm{GMO}}(M_\eta^2-M_\pi^2) \\
&+4\left( M_\pi ^4-4M_K^2M_\pi ^2+4M_\eta ^2M_\pi ^2-M_\eta ^4\right)L_5^r+24\left( M_\eta ^4-M_\pi ^4\right)(2L_7+L_8^r)\\
&
\hspace{-1cm}
+\frac{\left(40M_K^6+8\left(M_\eta ^2-10M_\pi ^2\right)M_K^4 -3M_\pi^2(M_\pi^2+M_\eta^2)^2+2M_K^2(17M_\pi^4+25M_\pi^2M_\eta^2-6M_\eta^4)\right)F_\pi ^2 \mu _K}{6\left( M_K^2-M_\pi
^2\right) } \\
&+\frac{\left(-70M_K^4+2\left( 22M_\pi ^2+19M_\eta ^2\right) M_K^2-19M_\pi
^4+9M_\eta ^4-12M_\pi ^2M_\eta ^2\right) F_\pi ^2\mu _\pi }{6\left(
M_K^2-M_\pi ^2\right) } \\
&+\frac{\left(14M_K^4-M_\pi^4+21M_\eta ^4 +6M_\pi^2M_\eta^2 -2M_K^2\left(M_\pi^2+18M_\eta^2\right)\right)F_\pi ^2 \mu _\eta }{2\left(
M_K^2-M_\eta ^2\right) }\\
&+\frac{-408M_K^4+\left( 73M_\pi ^2+387M_\eta ^2\right) M_K^2+8M_\pi^4-63M_\eta^4 -M_\pi^2M_\eta ^2}{192\pi ^2}\,.
\end{split}
\end{align}
$\bullet\quad \pi K$
\begin{align}
\varphi _{\pi K}&=-4L_3+\frac{\left( M_\pi ^2-4M_K^2\right) F_\pi ^2\mu _K}{%
6M_K^2\left( M_K^2-M_\pi ^2\right) }+\frac{\left( M_\pi ^2-4M_K^2\right) F_\pi ^2\mu
_\pi}{12M_\pi ^2\left( M_K^2-M_\pi ^2\right) }-\frac{ F_\pi ^2\mu _\eta}{M_\pi ^2-M_\eta ^2}\,,
\\
\varepsilon _{\pi K}&=12(4L_2^r+L_3)-\frac{3F_\pi ^2\mu _K}{M_K^2-M_\pi ^2}+%
\frac{15F_\pi ^2\mu _\pi }{2(M_K^2-M_\pi^2)}+\frac{6F_\pi ^2\mu _\eta }{%
M_\pi ^2-M_\eta ^2}+\frac 1{64\pi ^2}\,,
\\
\begin{split}
\delta _{\pi K}&=24(4L_1^r+L_3)+\frac{3\left( 3M_\pi ^2-4M_K^2\right)F_\pi ^2
\mu _K}{2M_K^2\left( M_K^2-M_\pi ^2\right) }+\frac{3\left( 7M_\pi
^2-8M_K^2\right) F_\pi ^2\mu _\pi }{4M_\pi ^2\left( M_K^2-M_\pi ^2\right) }
\\ &
-\frac{3F_\pi ^2\mu _\eta }{M_\pi ^2-M_\eta ^2}-\frac{23}{64\pi ^2}\,,
\end{split}\\
\begin{split}
F_\pi ^2(\gamma_{\pi K}-1) &=8M_\pi ^2L_5^r-\frac{2M_K^2F_\pi ^2\mu _K}{M_K^2-M_\pi ^2}-\frac{\left( M_K^2-5M_\pi ^2\right)F_\pi ^2 \mu _\pi }{2(M_K^2-M_\pi^2)}\\ &+\frac{\left(2M_K^2+M_\pi ^2-3M_\eta ^2\right) F_\pi ^2\mu _\eta }{M_\pi ^2-M_\eta ^2}
-\frac{5M_K^2+M_\pi ^2}{%
192\pi ^2}\,,
\end{split}\\
\begin{split}
F_\pi ^2(\beta_{\pi K}-1) &=32\left( M_K^2+M_\pi ^2\right)L_4^r+8L_5^rM_\pi ^2-\frac{15M_\pi ^2+19M_K^2}{64\pi ^2F_\pi ^2}
\\ &
+\frac{\left(3M_\pi^4+3M_K^2M_\pi^2-8M_K^4\right)F_\pi ^2 \mu _K}{M_K^2(M_K^2-M_\pi ^2)} -\frac{\left(3M_\pi^4-3M_K^2M_\pi^2-4M_K^4 \right)F_\pi ^2 \mu _\pi }{2M_\pi^2(M_K^2-M_\pi^2)} \\
&-\frac{\left(2M_\pi ^4+9M_\eta ^4+\left(4M_K^2+M_\pi ^2\right) M_\eta^2\right)F_\pi ^2 \mu _\eta }{2M_\eta ^2\left( M_\pi ^2-M_\eta ^2\right) }\,,
\end{split}\\
\begin{split}
F_\pi ^2(\alpha _{\pi K}-1)M_\pi M_K &=16L_4^r\left( M_K^4-4M_\pi ^2M_K^2+M_\pi ^4\right)+4L_5^rM_\pi^2 \left( M_\pi ^2-5M_K^2\right)+48M_K^2M_\pi^2(2L_6^r+L_8^r)\\
&-\frac{21M_K^4-25M_\pi ^2M_K^2+21M_\pi ^4}{192\pi ^2}+\frac{\left(2M_K^6-4M_K^4M_\pi^2+6M_\pi^4M_K^2-3M_\pi^6\right) F_\pi ^2\mu_K}{2M_K^2(M_K^2-M_\pi^2)} \\
& -\frac{\left(8M_K^6-13M_K^4M_\pi^2+12M_\pi^4M_K^2-6M_\pi^6\right)F_\pi ^2 \mu _\pi}{4M_\pi^2(M_K^2-M_\pi^2)} \\
&
\hspace{-0.8cm}
\hspace{-1.5cm}-\frac{\left(17M_\pi^6-24M_K^2M_\pi^4+3\left(27M_\eta^2+7M_\pi^2-36M_K^2\right)M_\eta^4+\left(24M_K^4+12M_\pi^2M_K^2+M_\pi^4\right)
M_\eta^2\right)F_\pi^2\mu_\eta}{24M_\eta^2\left(M_\pi^2-M_\eta^2\right) }\,.
\end{split}
\end{align}
$\bullet\quad KK$
\begin{align}
\varphi_{KK}&=24 L_2^r-\frac{17 F_{\pi }^2 \mu _K}{4M_K^2}-\frac{F_{\pi }^2 \mu _{\pi }}{4M_{\pi }^2}-\frac{7}{64 \pi ^2}\,,
\\
\varepsilon _{KK}&=12(2L_2^r+L_3)-\frac{F_\pi ^2\mu _K}{4M_K^2}-\frac{F_\pi ^2}8\left(\frac 2{M_\pi ^2}-\frac{27}{M_K^2-M_\pi ^2}\right) \mu _\pi +\frac{9F_\pi ^2\mu _\eta }{2\left( M_\pi ^2-M_\eta ^2\right) }-\frac 1{64\pi ^2}\,,
\\
\begin{split}
\delta _{KK}&=12(4L_1^r+L_3)-\frac{11F_\pi ^2\mu _K}{4M_K^2}-\frac
{F_\pi ^2}{16}\left(\frac{10}{M_\pi ^2}+\frac{27}{M_K^2-M_\pi ^2}\right)
\mu _\pi
\\ &
-\frac 98F_\pi ^2\left(\frac 3{M_\eta ^2}+\frac 2{M_\pi ^2-M_\eta
^2}\right) \mu _\eta -\frac{29}{128\pi ^2}\,,
\end{split}\\
\begin{split}
F_\pi^2\gamma _{KK} &=16M_K^2L_4^r-\frac{7M_K^2}{64\pi^2}-\frac{3F_\pi^2\mu_K}2
-\frac{3M_K^2\left(2M_K^2+M_\pi^2\right)F_\pi^2\mu_\pi}{8M_\pi^2(M_K^2-M_\pi^2)}\\ &+\frac{\left(M_\pi^4-3M_\eta^4+2M_\pi^2M_\eta^2+3\left(M_\eta^2-3M_\pi^2\right)M_K^2\right)F_\pi^2 \mu_\eta}{4M_\eta^2\left(M_\pi^2-M_\eta^2\right) }\,,
\end{split}\\
\begin{split}
F_\pi ^2(\beta _{KK}-1) &=16M_K^2L_4^r+8L_5^rM_\pi ^2-\frac{9M_K^2}{64\pi ^2}-\frac{3F_\pi ^2\mu_K}2-\frac{\left(6M_K^4+11M_\pi ^2M_K^2-20M_\pi ^4\right)F_\pi ^2 \mu _\pi }{%
8M_\pi^2(M_K^2-M_\pi^2)} \\
&+\frac{\left(M_\pi^4-9M_\eta^4+8M_\pi^2M_\eta^2+M_K^2\left(11M_\eta^2-9M_\pi ^2\right)\right) F_\pi ^2\mu_\eta }{4M_\eta ^2\left(M_\pi^2-M_\eta^2\right) }\,,
\end{split}\\
\begin{split}
F_\pi ^2(\alpha _{KK}-1)M_K^2 &=16M_K^4(-2L_4^r+6L_6^r+3L_8^r)+8L_5^rM_K^2\left( M_\pi ^2-3M_K^2\right)\\
&-\frac{7M_K^4}{192\pi^2}-\frac{3M_K^2F_\pi ^2\mu_K}2
-\frac{M_K^2\left( 6M_K^4+17M_\pi ^2M_K^2-20M_\pi ^4\right)F_\pi ^2 \mu _\pi }{%
8M_\pi^2(M_K^2-M_\pi^2)} \\
&
\hspace{-1.5cm}
-\frac{\left(M_\pi^6+2M_\eta^2M_\pi^4-3M_\eta^4M_\pi^2+3M_K^4(9M_\pi^2-7M_\eta^2) -6M_K^2\left(2M_\pi^4+5M_\eta^2M_\pi^2-7M_\eta^4\right)\right) F_\pi^2\mu_\eta}{12M_\eta^2\left(M_\eta^2-M_\pi^2\right) }\,.
\end{split}
\end{align}
}
In \cite{Spanele} they believe $K^+K^-\rightarrow K^+ K^-$ to be independent on $K^+K^-\rightarrow K^0\Ka$, but we know that isospin structure (Fierz-like identities) and the crossing symmetry dictate the relation between these two processes given by (\ref{stejna K}) together with (\ref{lisici se K}). Therefore, we have used the values of our parameters obtained from their $K^+K^-\rightarrow K^0\Ka$ amplitude to explicitly check their $K^+K^-\rightarrow K^+ K^-$ result $T_{ch}$.

\section{Short comment on the assumptions of the analyticity of the amplitude and the dispersion
relations}\label{disperzni diskuze}

The most important assumptions of the theorem -- the existence of
dispersion relations and the analyticity of the amplitude and its
absorptive parts are results of a complicated theory of analytic
properties of scattering amplitudes, which is even older than QCD
itself. A good and still valid summary of it is the article by Sommer
\cite{Sommer}.

We will not address this theory in more detail, just summarising
results interesting for us (details can be found in \cite{Sommer}
and \cite{moje}).

From the principles of axiomatic field theory (even without use of the unitarity of
S-matrix), Lehmann has proven that the amplitude with $s$ fixed at some physical value is
holomorphic in some finite region of the $u$-plane (if $s$ is above
the physical threshold, this region is the so-called small Lehmann
ellipse). Absorptive parts of amplitudes ($\im{A(s,u)}$ for
$s\ge\Sigma$ and similarly for $t$) are holomorphic in the large
Lehmann ellipses in the $u$-plane (depending on $s$) and there this
absorptive part has also a convergent partial wave decomposition.
Further, the $N$-times subtracted $u$-fixed dispersion relations can
be proven on the intersection of those large Lehmann ellipses for
$s\ge\Sigma$ and $t\ge\tau$ (if not empty). Since its semiminor
axis tends to zero for $s\rightarrow\infty$, this intersection is
just an interval on the negative real $u$-axis.

Taking into account the unitarity, Martin and others have succeeded in
enlarging the validity of the dispersion relations into the circle
$|u|<R$ with some fixed radius $R$.

There exist further methods to enlarge the region of validity as
well as the analyticity domain. Using the ones given in
\cite{Sommer} we end up with the results of Table~IV there. We are
interested in their specific application for our processes. If we
assume the isospin conservation and that all the mesons $\pi, K,
\eta$ are thereby stable (and forget about resonances) we get the
validity of fixed $u$ dispersion relations within the regions
depicted in Fig.~1 (the intersections of Lehmann ellipses are
marked by bold lines there). These regions are glued ellipses and have
been obtained analytically in \cite{moje} using procedures from
\cite{Sommer}. Let us remind that these regions are only the minimal
domains, where the dispersive relations are valid, proven directly
from the axiomatic theory. Using further methods we could also
extend these regions - e.g.\ we have not taken into account the
further specific crossing (and Bose) symmetries of some of the
amplitudes. Finally let us once more emphasise that the described
method is ineffective if we allow e.g.\ the $\eta\rightarrow3\pi$
decay and such cases have to be proven for unphysical masses of $\eta$ and then analytically continued in them.

Another assumption of our theorem was the existence of a point $s$, where the amplitude is analytic with respect to
all the values of variable $u$, for which the theorem should be valid - but in all the cases one can show that there exists a value $s$ for
which the small Lehmann ellipse is larger than the regions from the Fig.~1 and so we can conclude that both the assumptions
of the theorem are valid within these regions.

\begin{figure}
     \centering
     \subfigure[$\pi\pi\rightarrow\pi\pi$]{
          \includegraphics[width=.28\textwidth,height=.29\textwidth]{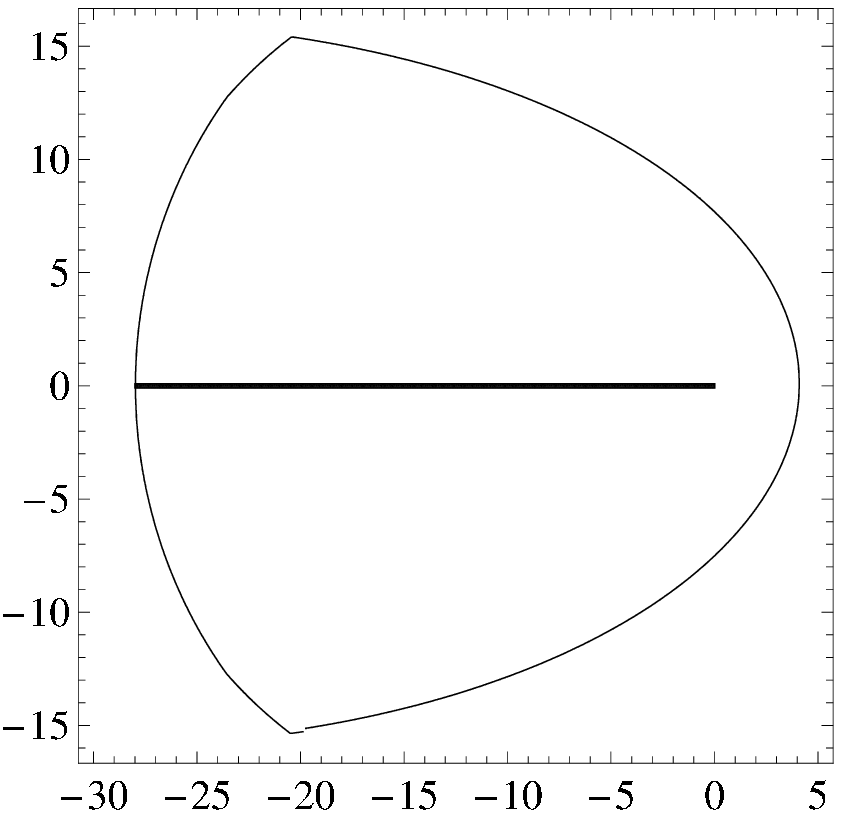}}
     \hspace{.1in}
     \subfigure[$KK\rightarrow KK$]{
          \includegraphics[width=.28\textwidth,height=.29\textwidth]{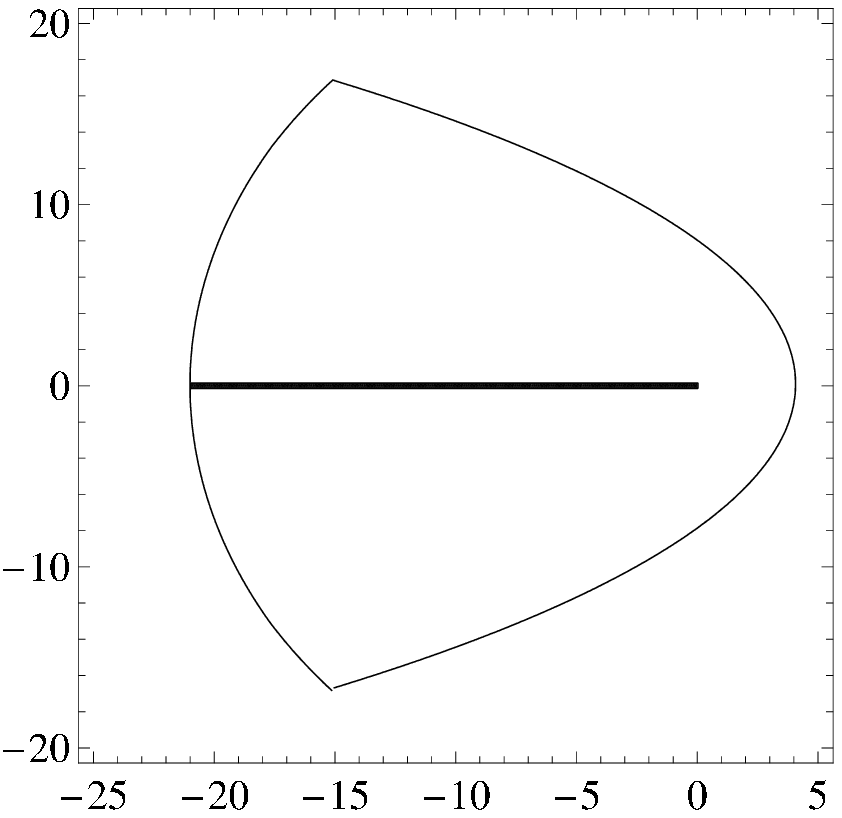}}
     \hspace{.1in}
     \subfigure[$\eta\eta\rightarrow\eta\eta$]{
           \includegraphics[width=.28\textwidth,height=.29\textwidth]
                {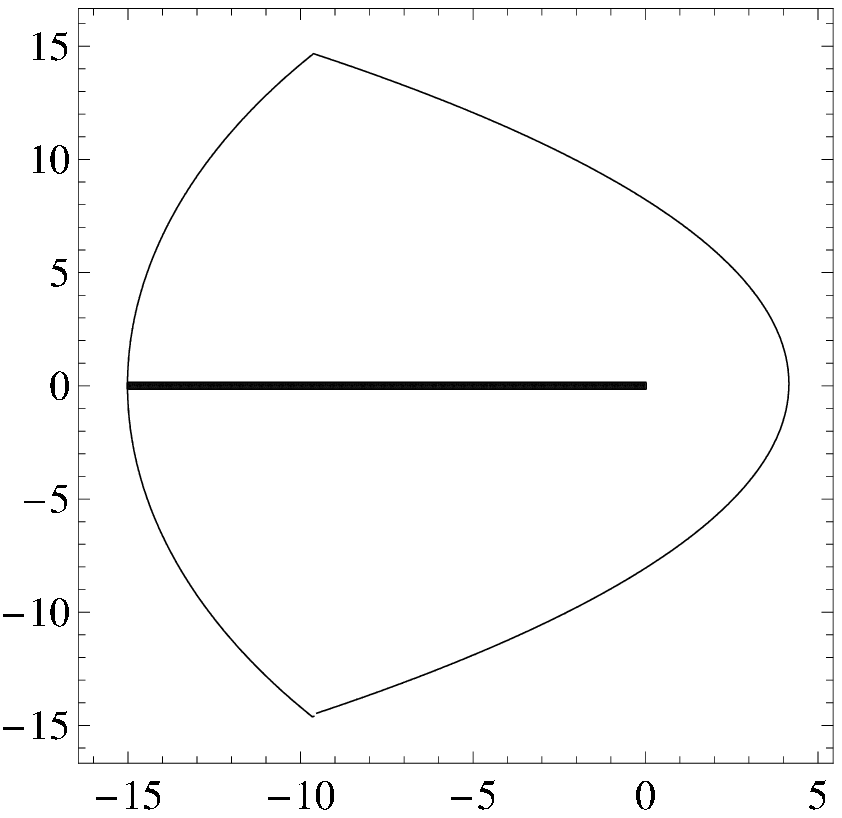}}\\
     \vspace{.05in}
     \subfigure[$\pi K\rightarrow K\pi$]{
          \includegraphics[width=.28\textwidth,height=.29\textwidth]{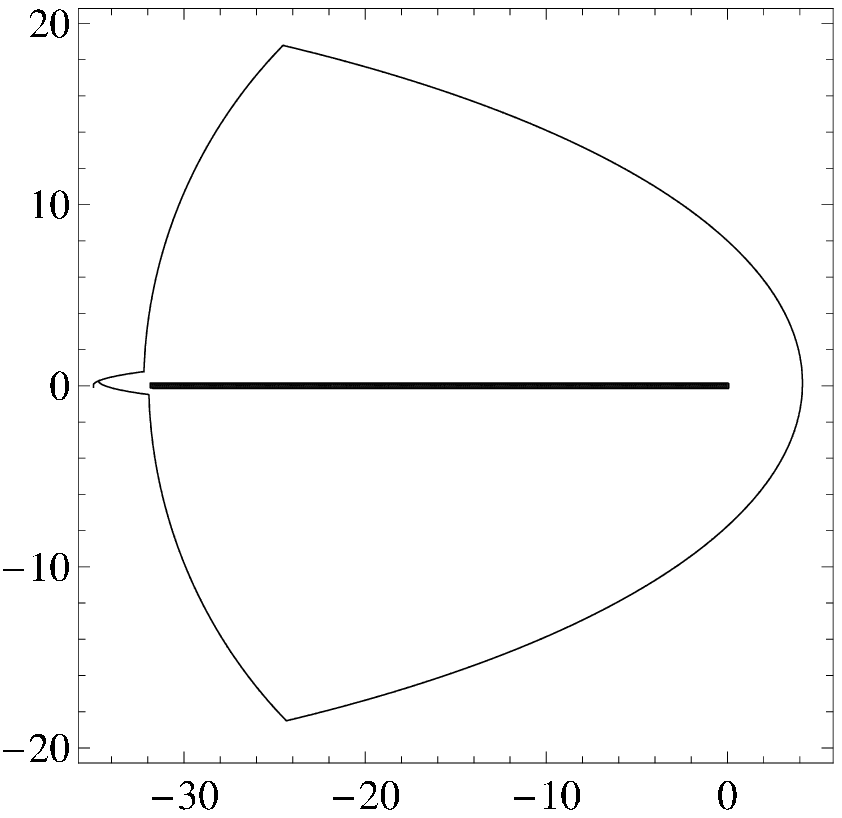}}
     \hspace{.1in}
     \subfigure[$\pi\eta\rightarrow\eta\pi$]{
           \includegraphics[width=.28\textwidth,height=.29\textwidth]
                {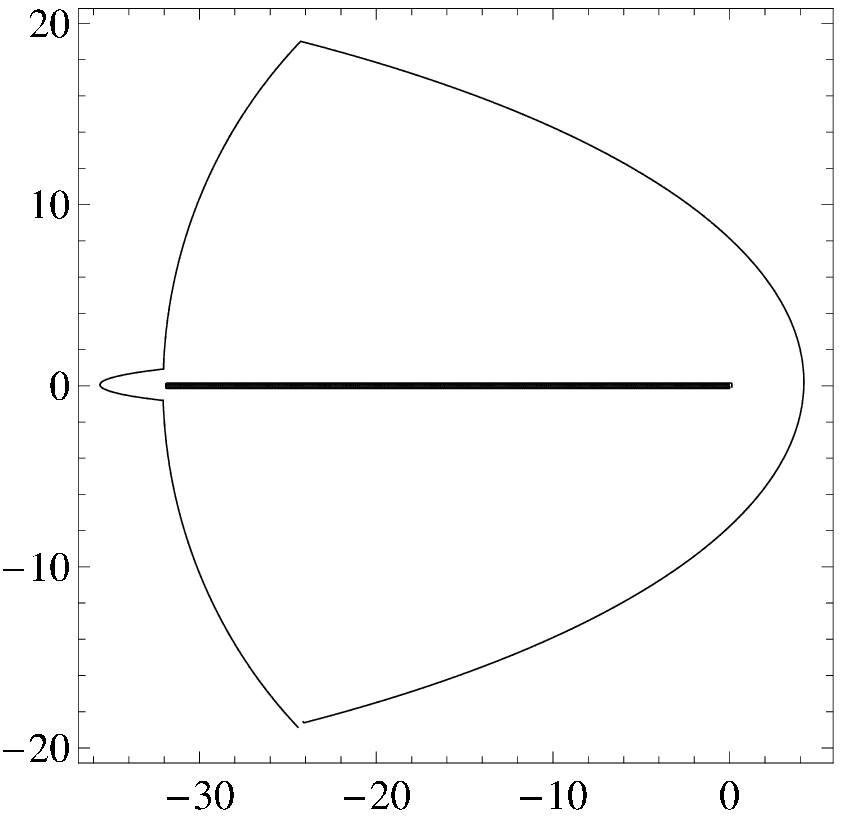}}
     \hspace{.1in}
     \subfigure[$K\eta\rightarrow\eta K$]{
          \includegraphics[width=.28\textwidth,height=.29\textwidth]{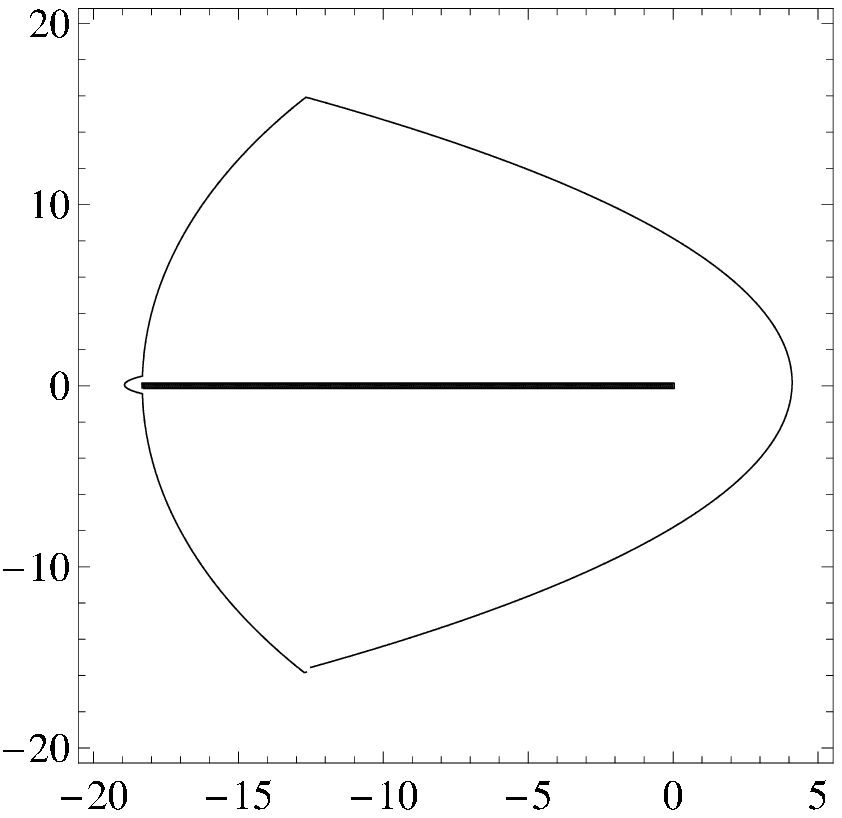}}\\
\vspace{.05in}
     \subfigure[$\pi\pi\!\rightarrow\!KK$ \& $\pi K\!\rightarrow\!\pi K$]{
          \includegraphics[width=.28\textwidth,height=.29\textwidth]{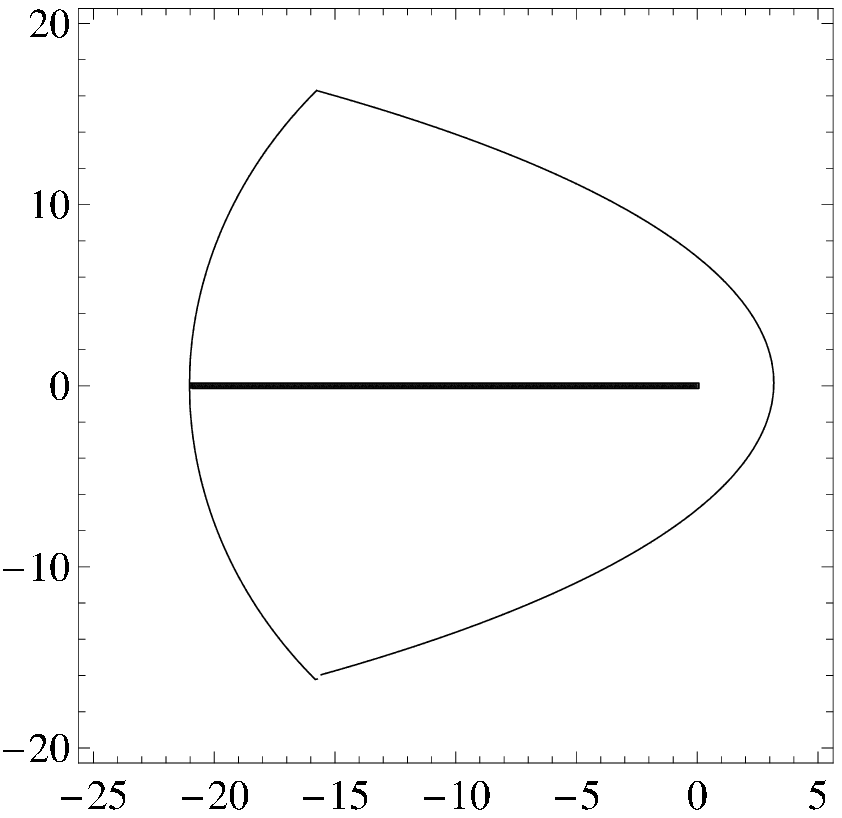}}
     \hspace{.1in}
     \subfigure[$\pi\pi\rightarrow\eta\eta$ \& $\pi\eta\rightarrow\pi\eta$]{
           \includegraphics[width=.28\textwidth,height=.29\textwidth]
                {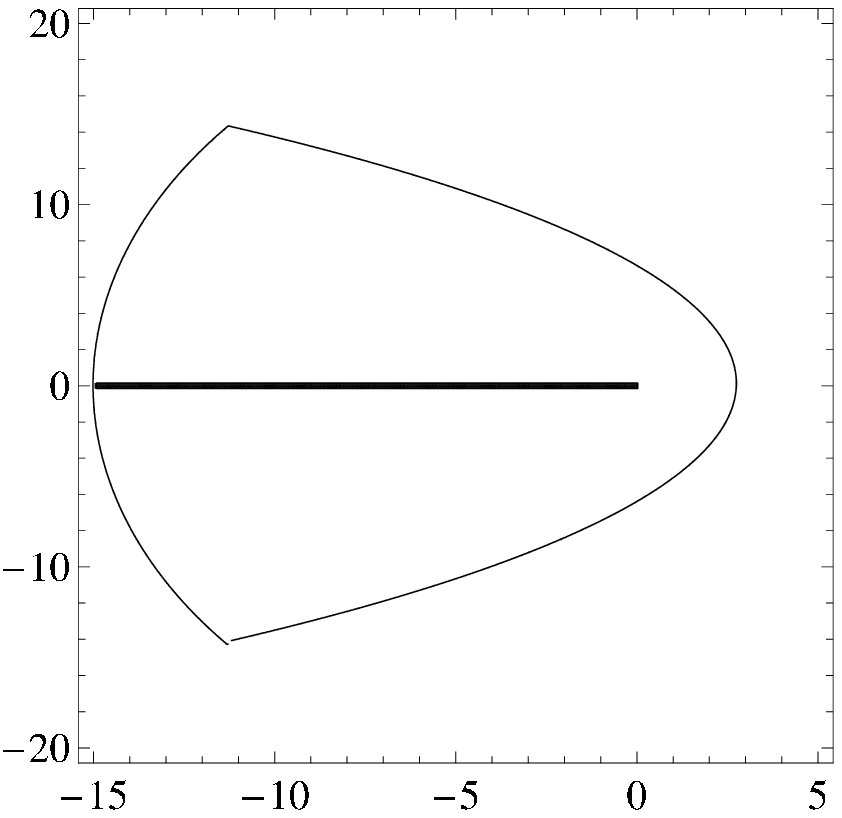}}
     \hspace{.1in}
     \subfigure[$KK\rightarrow\eta\eta$ \& $K\eta\rightarrow K\eta$]{
          \includegraphics[width=.28\textwidth,height=.29\textwidth]{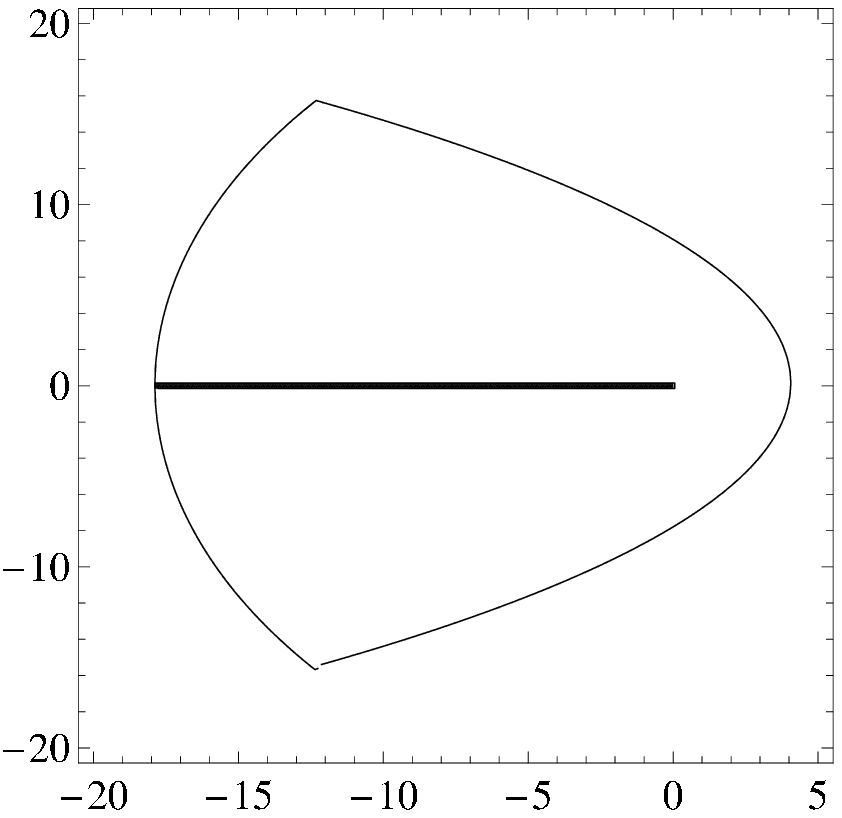}}\\
\vspace{.05in}
     \subfigure[$KK\rightarrow\pi\eta$ \& $K\pi\rightarrow K\eta$]{
          \includegraphics[width=.28\textwidth,height=.29\textwidth]{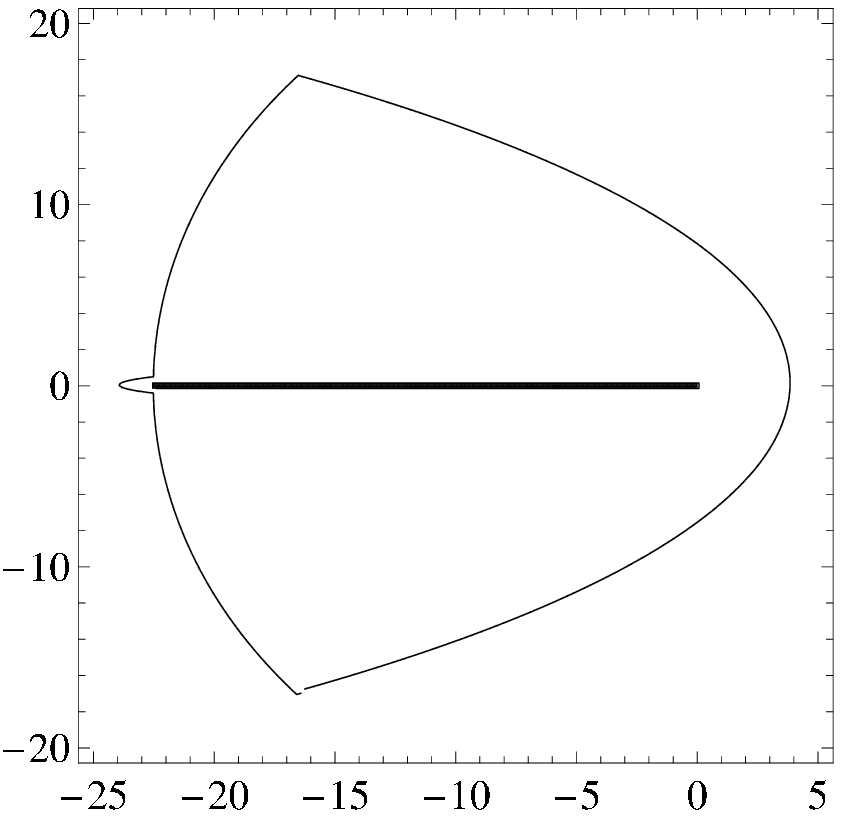}}
     \hspace{.1in}
     \subfigure[$K\pi\rightarrow\eta K$]{
          \includegraphics[width=.28\textwidth,height=.29\textwidth]{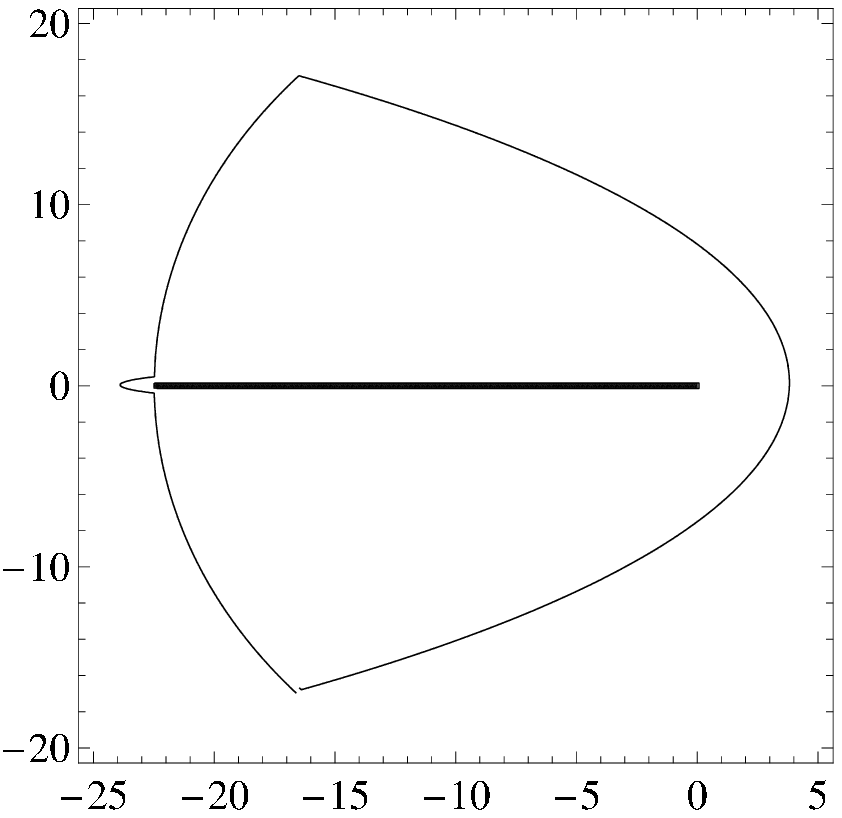}}

     \caption{The regions of validity of fixed $u$ dispersion relations (complex $u$-plane) in the multiples of $m_\pi^2$ and therefore also regions where the assumptions of our theorem are valid. With the bold lines, the intervals of validity coming from the use of just the Lehmann theory (ellipses) are depicted.}
\end{figure}

\end{document}